\documentclass[pdflatex,sn-mathphys-num]{sn-jnl}
\usepackage{graphicx}%
\usepackage{multirow}%
\usepackage{amsmath,amssymb,amsfonts}%
\usepackage{amsthm}%
\usepackage{mathrsfs}%
\usepackage[title]{appendix}%
\usepackage{xcolor}%
\usepackage{textcomp}%
\usepackage{manyfoot}%
\usepackage{booktabs}%
\usepackage{algorithm}%
\usepackage{algorithmicx}%
\usepackage{algpseudocode}%
\usepackage{listings}%
\usepackage{float}%

\usepackage{subcaption}
\usepackage{makecell}
\usepackage{hyperref}
\usepackage{footmisc}
\usepackage{longtable}
\usepackage[tableposition=top]{caption}
\usepackage{pdflscape}
\usepackage{url}

\newcommand{\RMunits}{rad$\,$m$^{-2}\,$}

\newcommand{\DMunits}{pc\,cm$^{-3}\,$}
\newcommand{\bursth}{bursts hr$^{-1}\,$}
\newcommand{\Qf}{Q_{\rm favg}}
\newcommand{\Uf}{U_{\rm favg}}
\newcommand{\Vf}{V_{\rm favg}}
\newcommand{\totbursts}{5526\,} 
\newcommand{\burstrate}{35.9$\pm$0.6\,}
\newcommand{\tottime}{$\sim$154\,}
\newcommand{\noisefreq}{11.1231\,Hz\,}
\newcommand{\firstobs}{60342}
\newcommand{\lastobs}{60827}
\newcommand{\SNfrac}{$\sim$5\%\,}
\newcommand{\avgRM}{328.0$\pm$0.1\,}
\newcommand{\avgDM}{528.78$\pm$0.4\,}
\newcommand{\PAtimebin}{65.5 ms\,}

\hypersetup{
    colorlinks=true,
    linkcolor=blue,
    filecolor=magenta,      
    urlcolor=magenta,
    pdfpagemode=FullScreen,
    }
\def\code#1{\texttt{#1}}

\begin{document}

\title[A fast radio burst cyclone in technicolour]{A fast radio burst cyclone in technicolour: evidence of plasma lensing}


\author*[1]{\fnm{Pavan A.} \sur{Uttarkar}}\email{puttarkar@swin.edu.au }

\author[1,2]{\fnm{Ryan~M.} \sur{Shannon}}\email{rshannon@swin.edu.au}

\author[3]{\fnm{Kelly} \sur{Gourdji}}\email{Kelly.Gourdji@csiro.au}

\author[1,2]{\fnm{Adam~T.} \sur{Deller}}\email{adeller@astro.swin.edu}

\author[4]{\fnm{Pravir} \sur{Kumar}}\email{pravir.kumar@weizmann.ac.il}

\author[5,6]{\fnm{Navin} \sur{Sridhar}}\email{navinsridhar@stanford.edu}

\author[1,2]{\fnm{Marcus~E.} \sur{Lower}}\email{mlower@swin.edu.au}

\author[7]{\fnm{Artem} \sur{Tuntsov}}\email{Artem.Tuntsov@manlyastrophysics.org}

\author[1,2]{\fnm{Atharva~D.} \sur{Kulkarni}}\email{adkulkarni@swin.edu.au}

\author[2,8]{\fnm{Simon C.-C.} \sur{Ho}}\email{simon.ho@anu.edu.au}

\author[1,2]{\fnm{Yuanming} \sur{Wang}}\email{yuanmingwang@swin.edu.au}

\author[1,2]{\fnm{Joscha} \sur{N.\ Jahns-Schindler}}\email{jjahnsschindler@swin.edu.au}


\affil*[1]{\orgdiv{Centre for Astrophysics and Supercomputing}, \orgname{Swinburne University of Technology}, \orgaddress{ \city{Hawthorn}, \postcode{3122}, \state{VIC}, \country{Australia}}}

\affil[2]{\orgname{ARC Centre of Excellence for Gravitational Wave Discovery  (OzGrav)}, \orgaddress{\city{Hawthorn}, \postcode{3122}, \state{VIC}, \country{Australia}}}

\affil[3]{\orgname{Australia Telescope National Facility}, \orgname{CSIRO, Space and Astronomy}, \orgaddress{\street{PO Box 76}, \city{Epping}, \postcode{1710}, \state{NSW}, \country{Australia}}}

\affil[4]{\orgdiv{Department of Particle Physics and Astrophysics}, \orgname{Weizmann Institute of Science}, \orgaddress{\street{Herzl St}, \city{Rehovot}, \postcode{76100}, \country{Israel}}}

\affil[5]{\orgname{Department of Physics}, \orgname{Stanford University}, \orgaddress{\street{382 Via Pueblo Mall}, \city{Stanford}, \state{CA}, \postcode{94305}, \country{USA}}}

\affil[6]{\orgname{Kavli Institute for Particle Astrophysics \& Cosmology}, \orgname{Stanford University}, \orgaddress{\street{452 Lomita Mall}, \city{Stanford}, \state{CA}, \postcode{94305}, \country{USA}}}

\affil[7]{\orgname{Manly Astrophysics}, \orgaddress{\street{41-42 East Esplanade}, \city{Manly}, \postcode{2095}, \state{NSW}, \country{Australia}}}

\affil[8]{\orgname{Research School of Astronomy and Astrophysics, The Australian National University}, \orgaddress{\city{Canberra}, \postcode{2611}, \state{ACT}, \country{Australia}}}


\abstract{Fast radio bursts (FRBs) are bright, energetic, radio pulses of extragalactic origin \cite[][]{Lorimer:2007,Tendulkar:2017, Bannister:2019}. A dichotomy has emerged in the population: some produce repeat bursts \cite[][]{Spitler:2016}, while the majority do not \cite[][]{Shannon:2018,Chime/FRBCollaboration2026}. Most repeating sources only show rare repetitions \cite[][]{Chime/FRBCollaboration2026}, and none have been studied extensively over the wide bandwidths necessary to disentangle the physical processes that produce emission from distortions to bursts caused by intervening ionised gas \cite[][]{Li:2021, Sridhar+21}. Here we present radio observations of the most active repeating source, FRB 20240114A. Using an ultrawideband receiving system \cite[][]{Hobbs:2019}, we have detected 5526 repetitions, revealing an extreme spectral and temporal variability in the burst emission. The bursts exhibit longer-term broadband variations in central emission frequency over multiple months, and narrowband bursts that have correlations in central frequencies on time scales of milliseconds to minutes. The spectral and temporal properties are consistent with the source undergoing magnification by foreground plasma lenses \cite[][]{Clegg:1998,Cordes:2017}, potentially embedded in a turbulent circumsource medium. This extreme example highlights the role of plasma lenses in the observed properties of burst emission and can explain the diversity in activity and energetics of the entire FRB population. }


\maketitle

\section*{Main}

Despite the detection of only three bursts in one week, the repeating fast radio burst  (FRB) source FRB~20240114A, discovered by the Canadian Hydrogen Intensity Mapping Experiment (CHIME) telescope \cite[][]{Shin:2024},  was speculated to be highly active because the transit telescope has a limited exposure time ($\sim$4 minutes per day) at the relatively southern declination of $\approx +4^\circ$ \citep{Shin:2025}. Since discovery, many telescopes have observed the source \cite[e.g.,][]{Joshi:2024,Limaye_ATEL:2024,Ould-Boukattine:2024,Zhang_ATEL:2024, Limaye:2025, Zhang:2025}, with some able to measure burst positions with sub-arcsecond precision sufficient to identify a host galaxy \citep{Snelders:2024}. Optical follow-up observations have since localised the source to a dwarf star-forming galaxy with a redshift of 0.1300$\pm$0.0002 \citep{Bhardwaj:2025}.

After the FRB discovery was announced on 26 January 2024, we initiated an observing campaign targeting the source using {\em Murriyang}, the 64-m Parkes radio telescope.  Our observations primarily used its ultra-wideband low (UWL, frequency range $704-4032$\,MHz) receiving system \citep[][]{Hobbs:2019}; however, on a few occasions,  we used the higher-frequency ($8-9$\,GHz) MARS receiver. The vast majority ($>95\%$ of the total observing time) of the observations were conducted with the UWL receiver, where we made all of our burst detections\footnote{We observed the source with the less sensitive MARS receiving system for 4\,hr and did not detect any bursts, see Section \ref{sec:observation} in the Methods}. The UWL system can observe a contiguous band of 6:1 fractional bandwidth, providing an unrivalled opportunity to study FRB emission properties and the passage of the emission through intervening plasma.  

Our follow-up observations show that within the UWL spectral band, FRB~20240114A is the most active FRB source known. Here, we report a high-cadence campaign lasting $\sim$16 months (see Section \ref{sec:observation} in the Methods for details). In  a total integration time of \tottime hours, we have detected \totbursts\,bursts above a signal-to-noise ratio (S/N) of $7.5$, corresponding to a burst rate of \burstrate \bursth. 
Our observations quintuple the number of FRBs detected by {\em Murriyang}  in its entirety of $\sim$ two  decades of searches \cite[e.g.,][]{Petroff:2016, Anna-Thomas:2023, Kumar:2023}, which includes one-off bursts, but is dominated by repetitions from known sources.  
The high activity of the source and the wide band of the observations have enabled us to investigate several outstanding questions about FRB emission and burst propagation through the surrounding complex magneto-ionic environments.

\begin{sidewaysfigure}
    \centering
    \includegraphics[width=1\linewidth]{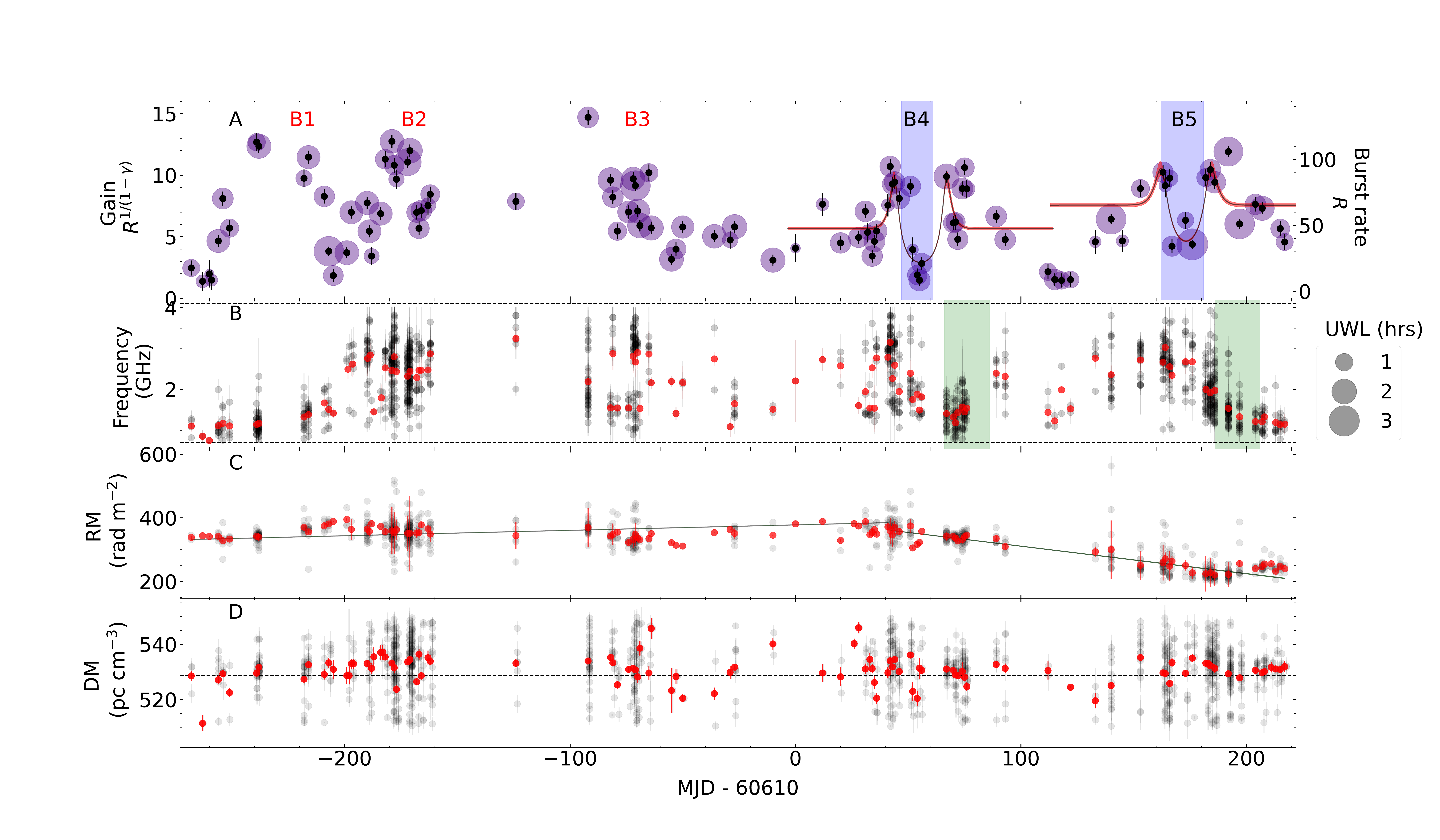}
    \caption{FRB~20240114A burst activity and spectral occupancy. Panel A: Temporal evolution of the gain, as estimated from the burst rate.
    The on-source integration time for the observations is shown as the purple shaded circle. The calculation of uncertainties for gain is discussed in Section \ref{sec:Physical_model} in the Methods. The ESE model fits to the burst storms B4 and B5 are shown in solid black lines. The uncertainties in the model fit are shown in red-shaded regions.
    Panel B: Spectral occupancy over the duration of observation for bursts that were detected with S/N $>$20, the grey points show the central frequency, and the vertical lines are the spectral occupancy. The red points show the mean spectral occupancy for each epoch.
    Panel C shows the RM of the source for all the bursts $>$20 S/N in black. Further discussion of the measurement of RM is presented in Section \protect\ref{sec:polarisation_measurment} in the Methods. 
    The red points show the average of all RMs across epochs. The uncertainties are calculated using the propagation of uncertainties. We show a straight-line fit to the evolution of RM over the course of the follow-up in a solid black line. The source showed a modest increase in RM of $\sim$0.2 day$^{-1}$\protect\RMunits until MJD $\sim$60684. However, the source shows an unambiguous decrease in the RM of $\sim$0.9 \RMunits day$^{-1}$ after MJD $\sim$60684. Panel D shows the DM values for bursts with S/N $>$20. See Section \ref{sec:DM_measurments} in the Methods for further discussion of DM measurements. The average DM (\avgDM) is shown as a black dashed line. 
   }
    \label{fig:burst_activity}
\end{sidewaysfigure}

We detect multiple episodes of increased activity, which we refer to as ``burst storms''. We empirically define burst storms as intervals where the burst rate exceeds 60 bursts hr$^{-1}$, following an established convention \citep{Nimmo:2023}. 
We identify five significant burst storms during our campaign, labelled B1 to B5 in Figure \ref{fig:burst_activity}.  
The burst rate range from as low as  2$\pm$1 bursts hr$^{-1}$ to a peak of 126$\pm$11 bursts hr$^{-1}$ during one of the storms.
Notably, we detect repetitions in every one of our UWL observations. Only FRB~20190520B has been observed to be similarly persistently active \citep{Niu2026}, albeit using the significantly more sensitive Five Hundred Metre Aperture Spherical Telescope (FAST).

We observe non-monotonic temporal variation of activity during burst storms. Notably, there is evidence for quenching of activity during the centre of the burst storm (like the eye of a cyclone), which we highlight in blue shading in Panel A of Figure \ref{fig:burst_activity}.
The burst storms arrive at relatively regular intervals, occurring every  $\sim$70 days, and last approximately half that time. 
While the burst storms are regular, there is no evidence for periodicity. 
We searched for periodicity using the fast folding algorithm \citep[FFA;][]{Staelin:1969} and Lomb-Scargle periodogram \citep{Lomb:1976, Scargle:1982} and did not detect significant periodicity with either method (see Methods Section \ref{sec:periodicity}).
This puts the source at odds with some other active repeating FRBs, which have been observed to have quasi-periodic activity cycles \cite[e.g.,][]{Rajwade:2020, Pleunis:2021, Pastor-Marazuela:2021}.

\begin{sidewaysfigure}    
    \centering
    \includegraphics[width=1.1\linewidth]{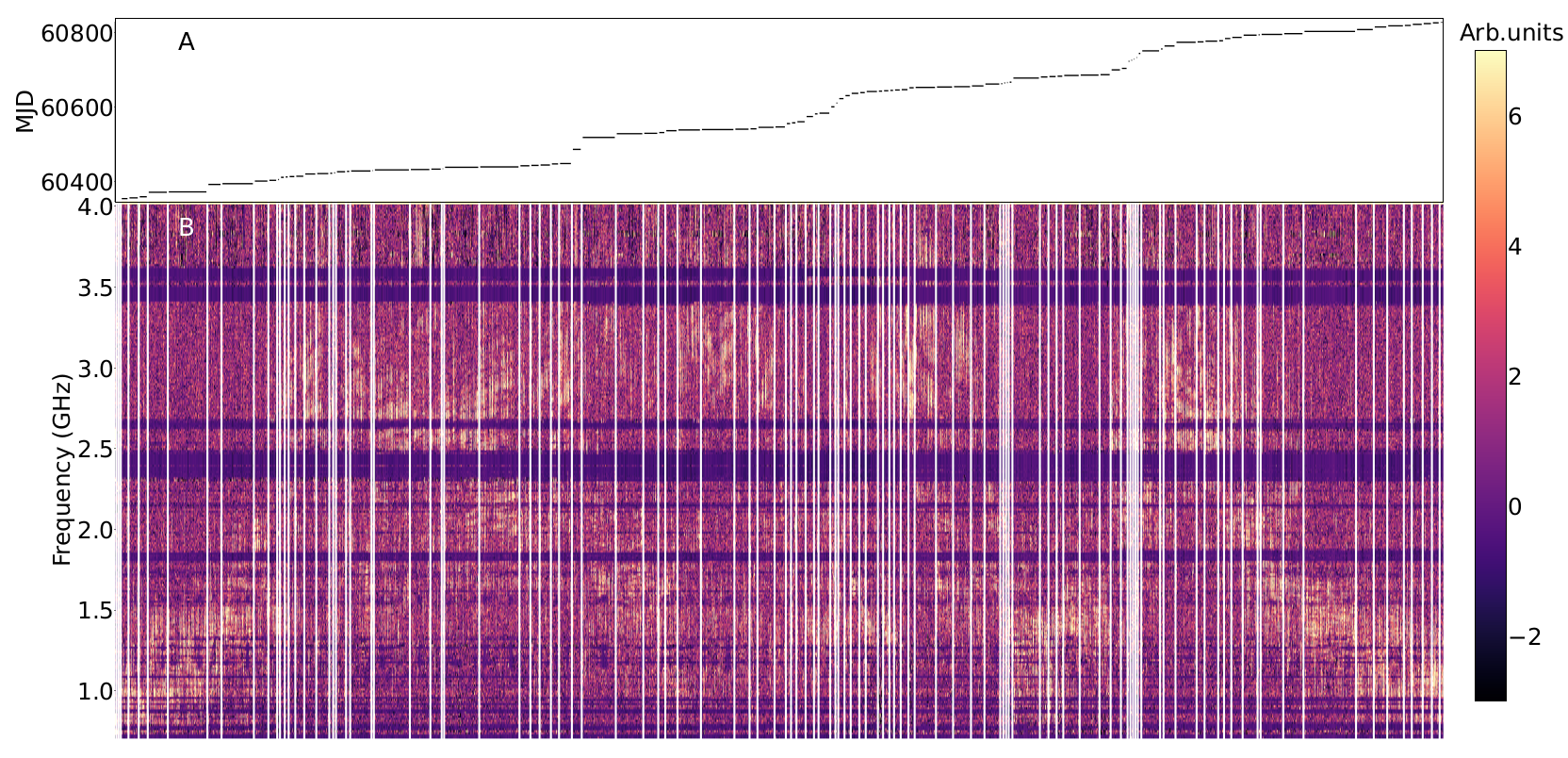}
    \caption{Burst spectral evolution. The bursts are displayed sequentially. Panel A shows the MJD for each burst. Panel B shows the dynamic spectra of individual bursts. Individual  bursts are integrated under the on-pulse region, with bursts averaged to a spectral resolution of $13$\,MHz. 
    Lighter colours denote higher flux density. Narrow  white spaces are used to delineate observing epochs. The spectral evolution can be clearly seen over the follow-up period. Initial epochs of follow-up of the source did not show any emission above 2 GHz. However, later epochs showed significant emission throughout the UWL band. The deep blue regions indicate the bands affected by interference.}
    \label{fig:ESE_evolution}
\end{sidewaysfigure}

The burst emission shows remarkable secular evolution in spectral activity. We present a dynamic spectrum of the bursts over the course of the campaign in Figure \ref{fig:ESE_evolution}. 
The dominant bands of burst emission slowly vary across the receiver band, in a way that has not been previously observed in any FRB. 
The bursts detected at the commencement of the observing campaign  and prior to MJD 60359 show no emission above $2$ GHz. However, after MJD 60372, emission is detected across the entire UWL bandwidth. 
The source shows a gradual evolution in the emission frequency during the burst storms until MJD 60431. 
During the well-monitored burst storms (labelled B4 and B5 in Figure \ref{fig:burst_activity}), the burst activity quenches at the peak of a storm. After the quenching, the source exhibits a dramatic decrease in the central frequency of the bursts. 
This effect is highlighted in the green shaded region of Panel B of Figure \ref{fig:burst_activity}. 
Repeating FRBs have been shown to exhibit preferred emission frequency extents that can change with time \citep{Gourdji:2019, Hessels:2019, Sheikh:2024}. However, no repeating source has been observed to exhibit such a remarkable evolution likely due to some combination of the lack of wide bandwidth observations, insufficient burst rates, and the absence of sustained long time-baseline follow-up campaigns.
Importantly, the spectral variability is unlike the chromatic burst activity observed in the nearby repeating source  FRB~20180916B \citep{Chime/FRBCollaboration:2020}, which is correlated with the phase within its periodic active window. This suggests that the cause of the spectral variability in  FRBs 20180916B and 20240114A could be different.

In addition to the correlated spectral activity over day-to-week timescales, the source exhibits short timescale (millisecond-minutes) spectral behaviour.
In this phenomenon, which we refer to as spectral memory, narrow-band bursts of similar shape are observed to occur over a nearly identical frequency range, but are separated by milliseconds to minutes. 
An example of two such bursts, which show the same bandwidth and central frequencies separated by $\sim$10 minutes, are shown in Extended Data Figure \ref{fig:narrow_carbon_copies}. Such sets of `carbon copies' are observed at multiple epochs, but at different centre frequencies. In some of the epochs, we detect bursts between carbon copies. However, they are not observed in the same frequency range. A complete listing of such events can be found in Extended Data Table \ref{tab:spec_mem}.
Such emission has not been previously observed in pulsars or FRBs.

The bulk foreground electron density, quantified by the dispersion measure (DM), and (density-weighted) net magnetisation, quantified by rotation measure (RM), show little to modest measurable variations. 
The first detected bursts by CHIME were reported to have a DM of  527.7 \DMunits and an RM of $\sim$325 \RMunits \citep{Shin:2025}.
Since then, the apparent DM has varied stochastically. Much of this variation can be attributed to complex burst spectro-temporal morphology, which can mimic DM variations. 
We observe stochastic variations in the RM within a given epoch. The average RM stays consistent within $\sim$3$\sigma$ until MJD 60684. However, we observe a steady decrease in the RM value of $\sim -0.9$ \RMunits day$^{-1}$ starting MJD $\sim$60684. 
The secular RM variation indicates the presence of structured magnetic fields in foreground plasma.  

The polarimetric properties of the individual bursts can also be used to study the burst emission mechanism and foreground propagation.  
We find that nearly all bursts exhibit a temporally constant polarisation position angle (PPA) over their duration. However, the PPA varies significantly between bursts. Extended Data Figure \ref{fig:PA_evolution} shows the PPA of all of the bright bursts. While bursts exhibit a wide distribution of PPA values, there are excesses at angles close to $-20^\circ$ and $+70^\circ$. 
We observe PPA variations in $\sim$ 15 bursts akin to the S-curve seen in pulsars in some of the bursts (see Extended Data Figure \ref{fig:PA_evolution}). 
We do not observe any long-term variations in the PPA of the bursts.
If bursts are produced in the magnetosphere of a neutron star, it suggests they originate at multiple locations within it.

\begin{sidewaysfigure}
        \centering
        \includegraphics[width=1\linewidth]{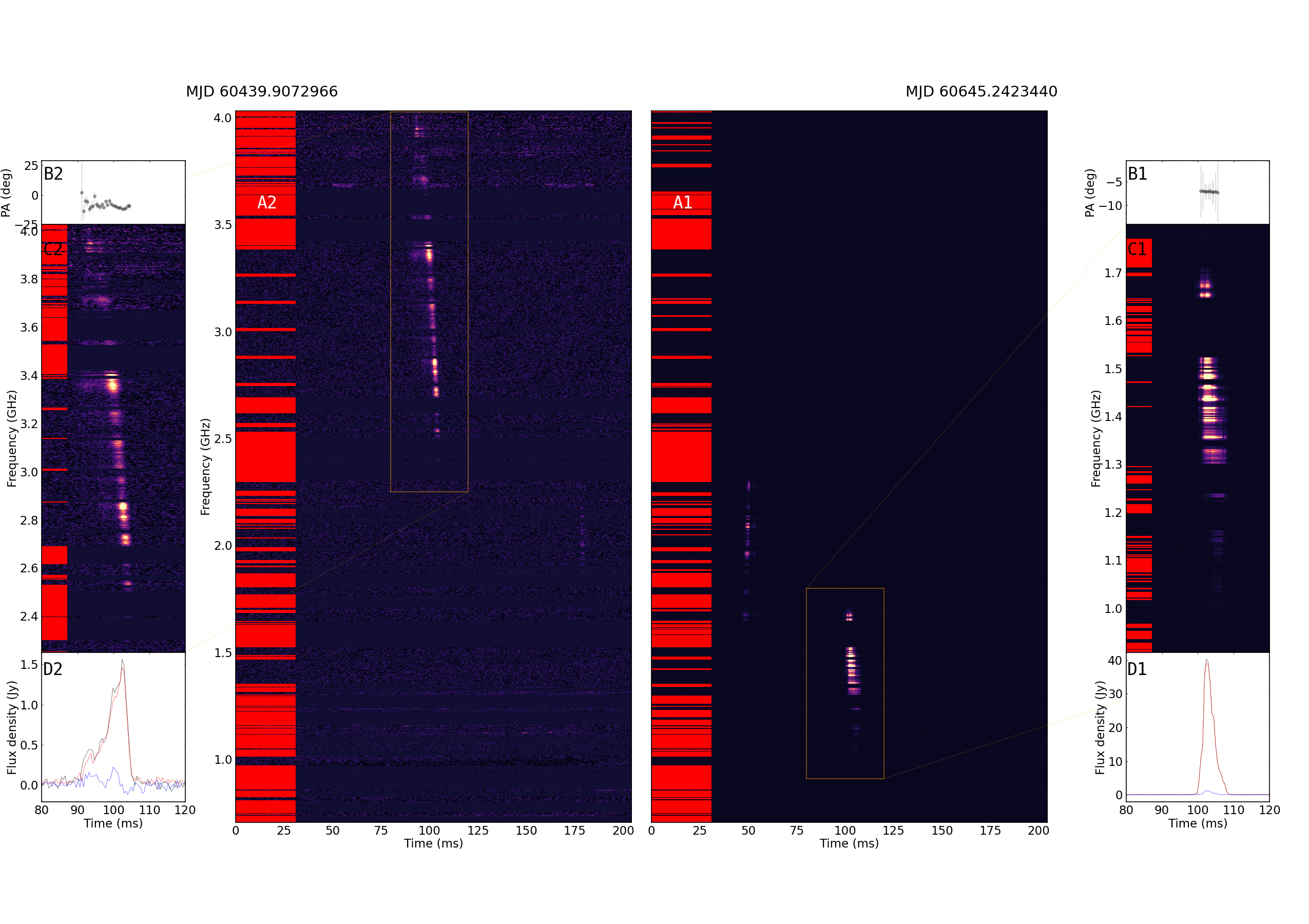}
        \caption{Bright bursts from FRB~20240114A. The bursts shown in Panels A1 and A2 were detected on MJDs 60439 and 60645, respectively. Bursts A1 and A2 were detected with an S/N of 155 and 338, respectively. We show the full-band dynamic spectra in the plot, which shows multiple bursts. The burst shown in Panel A2 is the brightest burst detected in our observation campaign.  The bursts show complex morphology with multiple components. 
        Panels B1 and B2 show the PPA angles for time bins with Stokes-I S/N $>$4.
        Panels C1 and C2   zoom in on the  spectral band in which the bright bursts are present. 
        The  frequency-averaged polarisation profile for both  bursts is shown in Panels D1 and D2. In these panels, the total intensity, linear, and circular polarisation profiles are shown in black, red, and blue lines, respectively. 
        The RFI masked channels are highlighted in red on the left-hand side for each channel. The times-of-arrival of bursts are shown at the top of the figure in MJD. 
    }
    \label{fig:bright_burst_B0}
\end{sidewaysfigure}

FRB~20240114A also produces some of the brightest emissions that have been observed from an extragalactic FRB.
We have observed bursts with fluences exceeding $\sim$100 Jy ms. 
In Figure \ref{fig:bright_burst_B0}, we show two of the most luminous bursts detected during the follow-up campaign. The bursts' spectral occupancy would have spanned the bandwidths of many FRB search systems \cite[e.g.,][]{2013Sci...341...53T,Shannon:2024}.
We estimate the S/N for these bursts to be 2548 and 130, corresponding to fluences of 125 Jy ms and 10 Jy ms for bursts A1 and A2, respectively.
Among \totbursts\, detected bursts, \SNfrac of bursts were detected above a detection S/N of 44.
The pulse width-fluence distribution of the source spans the distribution of both repeater population (e.g., 20121102A, 20190520B) detected using more sensitive telescopes (e.g., FAST, Arecibo) to the (apparent) non-repeater population (see Extended Data Figure \ref{fig:FRB20240114A_FLUENCE}). While some repeating FRB sources have been claimed to have a bimodal energy distribution \cite[e.g., FRBs 20121102A and 20201124A;][]{LiD:2021, Zhang:2021}, we do not observe evidence of such a bimodality for FRB~20240114A (see Extended Data Figure \ref{fig:FRB20240114A_FLUENCE}). However, it is possible that bimodality is present below our detection threshold.


There are several possible scenarios that explain the unprecedented properties of the burst emission, which implicate either the burst source itself or its environment.
We first consider models   that have been invoked to account for the chromaticity observed in FRB~20180916B -- the only other FRB that has been seen to have a clearly correlated frequency-dependent burst activity \citep{Pastor-Marazuela:2021, Pleunis:2021, Bethapudi:2023}. 
These models can be broadly bifurcated into two classes: extrinsic models, in which the engine is in a binary system, or intrinsic models, in which the emission mechanism itself produces frequency-dependent activity. The extrinsic models are motivated by Galactic analogs, such as the pulsar PSR B1259$−$63, in which a neutron star orbits and is eclipsed by a massive Be star, leading to systematic variations in the burst emission frequency due to free-free absorption \cite[e.g.,][]{Li:2021}.
Radius-to-frequency mapping, also observed in pulsars \citep{Cordes:1978}, in which lower frequencies are emitted higher up in the emission cone has been proposed as an intrinsic mechanism to explain the observed properties of FRB 20180916B. If the beam of the putative source slowly sweeps across the observer over multiple weeks, it would point to the presence of a very slowly rotating neutron star driving such emission. However, both cases predict  emission that would exhibit periodicity with greater significance and a more regular frequency-dependent activity. We observe neither of these properties.

Another possibility is that the bursts are produced by an accreting compact object with a jet that may wobble or precess due to variations in the confining pressure from the surrounding disk winds. In this scenario, FRB emission is expected to be produced along the cleaner jet funnel through either reconnection of the `striped' magnetic fields in the jet or through the synchrotron maser instability from a transverse, magnetised shock due to flares launched along the jet funnel \citep{Sridhar+21} (see Extended Data Figure \ref{fig:intrinsic_toy} in the Methods for a schematic model of this scenario). If we consider the maser emission, the corresponding peak frequency of the FRB scales as  $\nu_{\rm pk}(\theta)\propto\frac{L_{\rm q}^{3/4}\sigma_{\rm q}^{1/2}}{\Gamma_{\rm q}^2L_{\rm f}^{1/4}}$. Here, $\nu_{\rm pk}$ depends on the angle $\theta$ measured relative to the jet axis; $L_{\rm q}$, $\sigma_{\rm q}$, and $\Gamma_{\rm q}$ are the luminosity, plasma magnetisation, and Lorentz factor of the upstream, quiescent flow in the jet, and $L_{\rm f}$ is the luminosity of the flare that produces the shock. MHD simulations have motivated a structured jet profile, where $L_{\rm q}$ and $\Gamma_{\rm q}$ are expected to grow toward the jet edge, and the Poynting flux is concentrated in a hollow cone around the jet core \citep{Tchekhovskoy+08}. This means that higher frequency bursts would be emitted along the jet spine, and the lower frequency bursts (in the shaded green region in the second row of Figure \ref{fig:burst_activity}) would be emitted away from the jet spine. The variation in the burst rate could be either due to instabilities in the inner edge of the accretion disk leading to an enhanced flaring along the jet during burst storms, or a preferential orientation of the jet funnel with our line of sight. While the accreting jet model can explain variations in the spectral band of emission over multiple weeks and days, it cannot  explain the presence of carbon-copy bursts observed from FRB~20240114A.

The chromatic variability observed in the source emission is, however, consistent with being caused by extreme scattering events (ESEs). They have previously been observed in compact sources, namely radio-bright quasars \cite[e.g.,][]{Fiedler:1987, Bannister:2016} and pulsars \citep[e.g.,][]{Coles:2015, Kerr:2018}.  In these cases, they have been interpreted as the result of a divergent lens foreground to the source, produced by overdensities in the Milky Way interstellar medium. The lenses also produce frequency-dependent caustic magnification at their edges, with the spectral and temporal properties of the magnification depending on multiple physical parameters, including the geometry of the source and the density of the lensing plasma \citep{Clegg:1998, Cordes:2017}.
Magnification when the source is behind a caustic can increase background source brightness by orders of magnitude, especially for compact sources.  

Such lensing could also occur for FRBs. While lensing from the Milky Way could occur, it is also possible for lensing to occur in media closer to the FRB source. Indeed, many FRBs are observed to have dense and complex magneto-ionic environments that surround the emitting engine \citep[e.g.,][]{Feng:2022}.
Like a small subset of repeating FRBs \cite[][]{Chatterjee:2017,Niu:2022}, FRB~20240114A is associated with a persistent radio-continuum source \citep[PRS;][]{Bruni:2024}. The origin of PRSs has been speculated, among other models, to be powered by a young flaring magnetar in a supernova remnant \citep[e.g.,][]{Margalit:2018} or an accreting compact object \citep{Sridhar:2022}. This type of environment could naturally provide the source of plasma that forms lenses.

A model for plasma lensing has been developed in an attempt to explain some of the behaviour observed in the first repeating FRB \citep[][]{Cordes:2017}, which was an extension of a previous model for ESEs \cite[][]{Clegg:1998}. These models predict both the magnification (or gain) of a background source caused by a foreground lens and the presence of narrow-band spectral caustics; the latter of which we observe as carbon-copy bursts. 
We can use this model to explain the spectral and temporal properties of FRB~20240114A, with one modification. Given the intrinsic variability of the burst emission, it is difficult to directly infer lensing magnification from the flux density of the bursts. Instead, we estimate the magnification from the variability in burst rate. 
To do so, we first determine the intrinsic burst luminosity function and then estimate the magnification of the source from the change in the burst rate.  Details of the modelling are presented in Section \ref{sec:Physical_model} in the Methods.

\begin{figure}
    \centering
    \includegraphics[width=1.\linewidth]{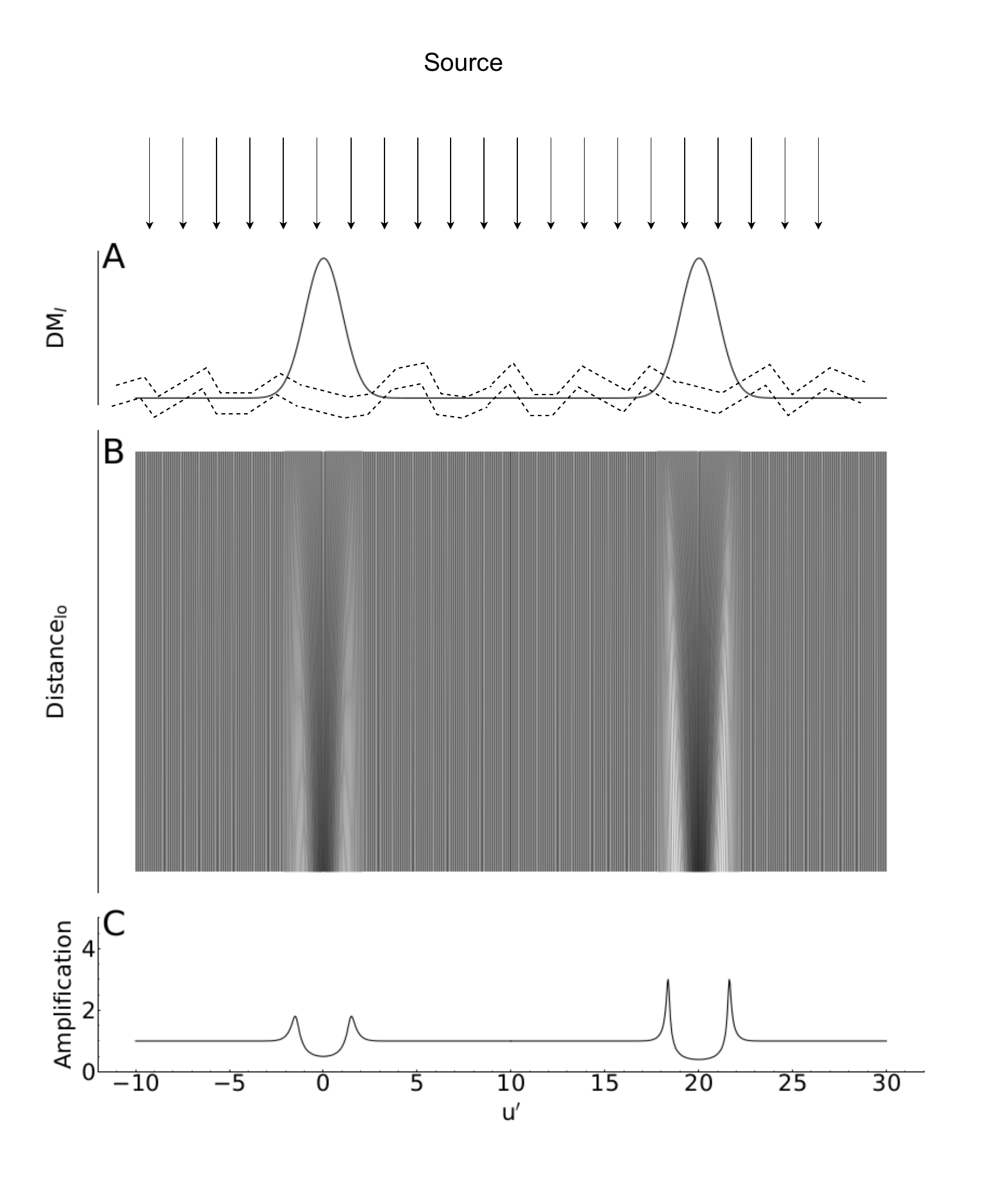}
    \caption{ESE Model for burst storms B4 and B5. Panel A shows the modelled plasma density variations. We use two Gaussian lenses to model burst storms B4 and B5. Panel B shows the ray trace diagram of the refracted waves due to the two lenses. Panel C shows the total effect of the lensing event, which causes an apparent change in the brightness of the source. The x-axis is the transverse scale.}
    \label{fig:ESE_toy}
\end{figure}

We apply this model to the burst storms B4 and B5, for which we have dense observing coverage. We fitted a four-parameter model \citep{Cordes:2017} for the gain. A model for this scenario is shown in Figure \ref{fig:ESE_toy}. 
We show the ESE model fitted to the data in Panel A of Figure \ref{fig:burst_activity}. The 1-$\sigma$ region is highlighted in red. The model well reproduces the observed burst rate variations. The model fit (the posterior parameters are shown in Extended Data Figure \ref{fig:posteriors_ESE}) includes a parameter $\alpha$ that characterises the lens scale. However, it is also degenerate with other parameters, namely the path DM associated with the lens, and the distances between the source, the lensing medium, and the observer. 
Of these, the most significant unknown is the distance from the source to the lens.

There are several possible locations for the lens, all within the FRB source's host galaxy. It is possible that the lens is embedded in a bow shock caused by the source moving through a dense medium. Another possibility is that the lensing  material resides in a supernova remnant surrounding the source. Finally, it is possible that the lens is in a larger H\,\textsc{ii} region along the line of sight. In the first two scenarios, the distances between the FRB source and lens will be of the order of $\sim$ parsecs. In the third case, we would expect the lens to be  at $\sim$ kiloparsec distances from the source \citep{Cordes:2017}.
If we assume the lens to be nearby to the FRB source (consistent with either the lens being associated with a magnetar wind nebula or supernova remnant and consistent with the upper limit  on the size of the PRS, which is smaller than $\sim$4 pc \citep{Bruni:2025}) and a DM contribution from the lens to be 4 \DMunits (based on the constraints on DM variations observed from this source over our follow-up campaign),
we estimate the lens size for the first and second burst storms to be 
$3.3\,{\rm AU}\,\left(\frac{\rm DM_{l}}{4 {\rm\,pc\,cm^{-3}}}\right)^{1/2}\,\left(\frac{d_{\rm sl}}{4\,{\rm pc}}\right)^{1/2}\,\left(\frac{\nu}{2.368\, {\rm GHz}}\right)^{-1}\,\left(\frac{\alpha}{0.9}\right)^{-1/2}$ 
and 
$4.0\,{\rm AU}\,\left(\frac{\rm DM_{l}}{4 {\rm\,pc\,cm^{-3}}}\right)^{1/2}\left(\frac{d_{\rm sl}}{4\,{\rm pc}}\right)^{1/2}\left(\frac{\nu}{2.368}\right)^{-1}\,\left(\frac{\alpha}{0.6}\right)^{-1/2}$, 
respectively. The lens properties are described in further detail in Section \ref{sec:Physical_model} in the Methods.
If the source was at kiloparsec distance from the lens, we expect the lens scale for B4 and B5 to be 52 AU and 64 AU, respectively.  

There is evidence for lensing events from circumsource material in Galactic pulsars. 
Some pulsars reside in young supernova remnants, such as the Crab pulsar, which  has been observed to undergo multiple plasma lensing events since its discovery \cite[e.g.,][]{Lyne:1975,Thierry:2025}. Interpreting a 1997 event in the Crab pulsar as plasma lensing, it was speculated that such events may be rare due to a paucity of plasma clumps \citep{Lyne:2001}. A larger number of plasma clumps in the FRB progenitor environment is required, given multiple such events over $\sim$16 months of follow-up. This would support a scenario in which FRB sources are embedded in the youngest supernova remnants and are the result of relatively recent supernova explosions.  Lensing has been observed in other pulsar systems too.  There is evidence of magnification of factors of $70-80$ due to ``extreme plasma lensing'' events in eclipsing pulsar PSR~B1957$+$20 \citep{Main:2018}. In this case, the lensing material is proposed to be the wind of the companion. The spectro-temporal variations  observed in FRB~20240114A could be due to a similar mechanism. It is plausible that such ESE signatures have been missed in other known repeating FRB sources, due to the large spectral coverage that is required to discern them from other physical processes.

The spectral properties observed in FRB~20240114A can be robustly accounted for by the ESE model, including the carbon copy behaviour, where the amplification of a band of frequencies might be leading to such emission, which changes with the structure of the plasma lenses along the path of the source and the observer. If plasma lensing were common for FRB sources, it could explain the variety of repetition rates, and, potentially, why some FRBs repeat and others do not.

\textbf{Author Contributions:} P.A.U. led the Murriyang data processing and drafting of manuscript with suggestions from A.T.D., A.K., K.G., M.E.L, N.S., P.K., R.M.S. S.C.-C.H., and Y.W.. S.C.-C.H., A.K., and Y.W. contributed to observation of the source with the Murriyang radio telescope. R.M.S. contributed in the interpretation of the data and data processing. P.A.U. is  PI of the Murriyang observing projects. A.T.D., K.G, M.E.L., P.K., R.M.S, and Y.W. are co-PIs for the Murriyang observations. P.K. developed the single pulse search pipeline used in the work. N.S and A.T. contributed to the  modelling of the source behaviour. S.C.-C.H. contributed to  data processing and verifying DM measurements.

\textbf{Funding:} P.A.U. and Y.W. acknowledge support through ARC Future Fellowship FT190100155. A.T.D., R.M.S., and J.J.-S. acknowledge support through ARC Discovery Project DP220102305. N.S. acknowledges support from the Simons Foundation (grant MP-SCMPS-00001470).

\textbf{Acknowledgement:} Murriyang, CSIRO’s Parkes radio telescope, is part of the Australia Telescope National Facility (https://ror.org/05qajvd42) which is funded by the Australian Government for operation as a National Facility managed by CSIRO. We acknowledge the Wiradjuri people as the Traditional Owners of the Observatory site. P.A.U. acknowledges helpful discussions with Ron Ekers, Keith Bannister, and Aditi Kaushik. The authors thank Marcin Glowacki for their comments on the manuscript. This work was performed on the OzSTAR national facility at Swinburne University of Technology. The OzSTAR program receives funding in part from the Astronomy National Collaborative Research Infrastructure Strategy (NCRIS) allocation provided by the Australian Government, and from the Victorian Higher Education State Investment Fund (VHESIF) provided by the Victorian Government.

\textbf{Competing interests:} The authors declare no competing interests.

\textbf{Data availability:} All the raw data will made available on the CSIRO's Data Access Portal (DAP; \url{http://data.csiro.au/}). The code base used for processing will be made available at \url{http://github.com/puttarkar/FRB20240114A_codebase}. The data for Figure \ref{fig:burst_activity} are available at \url{https://github.com/pavanuttarkar/FRB20240114A/blob/main/FRB20240114A_20sig_properties.pdf}.

\setcounter{figure}{0}
\captionsetup[figure]{name={\bf Extended Data Figure}}
\setcounter{table}{0}
\captionsetup[table]{name={\bf Extended Data Table}}


\clearpage

\section{Methods}\label{sec11}



\subsection{Murriyang observations}
\label{sec:observation}

We observed FRB~20240114A with the Murriyang radio telescope using both the ultra-wideband low receiver \citep[UWL;][]{Hobbs:2019}, which observes in a band spanning 704-4032\,MHz, and the higher frequency MARS receiver, which observes in a  band spanning 7881$-$8905 GHz. The observations presented here occurred  between MJDs \firstobs\,and \lastobs\,under project codes PX127 and P1338. All observations were made in pulsar search mode with temporal and spectral resolutions of 64 $\mu$sec and 0.5 MHz, respectively, and recorded with four coherency products enabling polarimetric studies. To mitigate intra-channel dispersion smearing, all the observations were coherently dedispersed  at a DM of 527.7 \DMunits. We used noise diode injection at the start of each observation for polarisation calibration, in particular to account for differential gain and phase between the two orthogonally polarised receptors of both systems. The noise diode injection was performed at a frequency of \noisefreq. The observations were recorded with the Medusa digital backend \citep{Hobbs:2019} until April 2025 and using the \code{Apollo} digital backend after the decommissioning of the former.

\begin{figure*}
    \includegraphics[width=1\columnwidth]{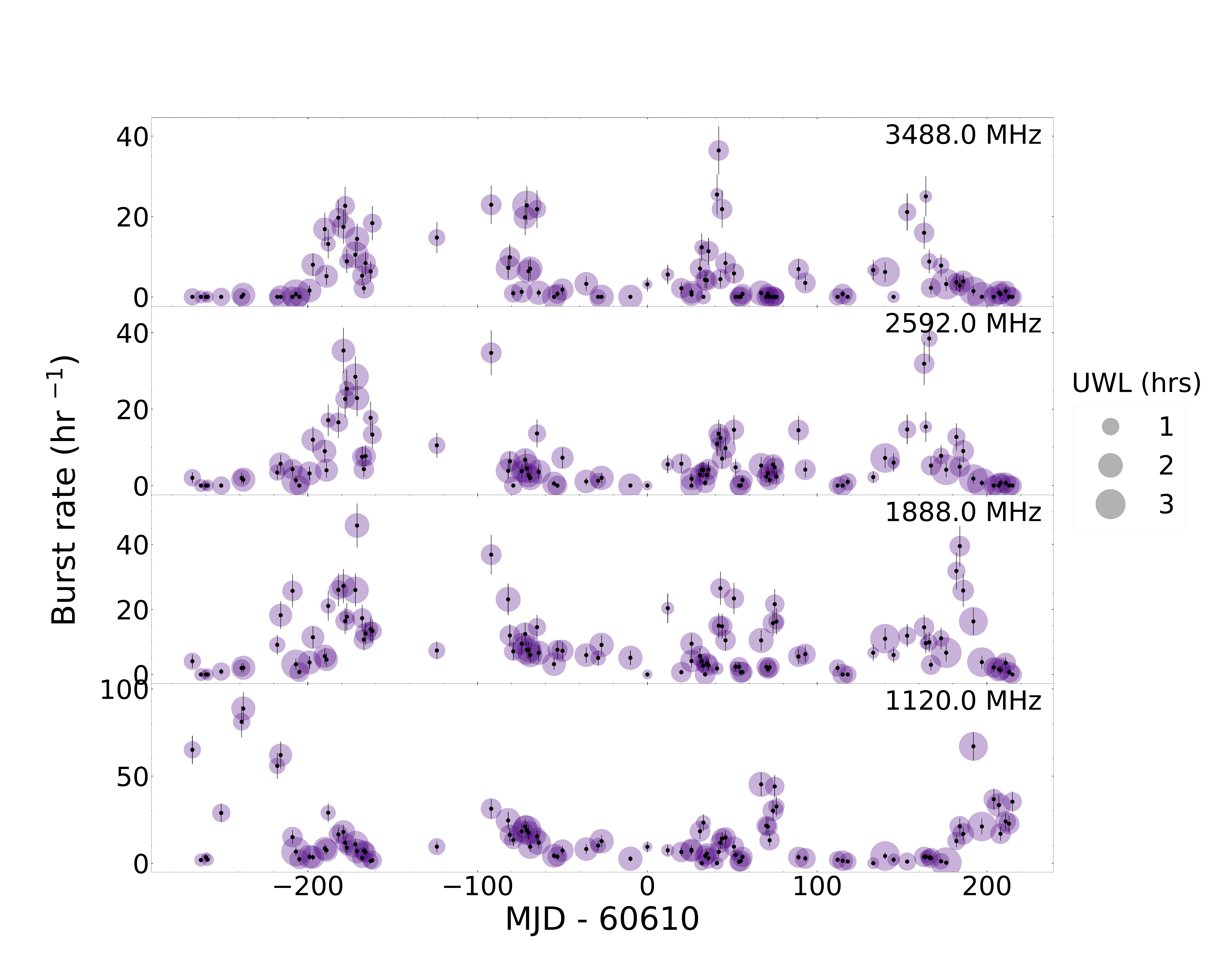}
    \caption{Subbanded burst rate for FRB~20240114A. The burst rate for four subbands is shown in separate panels. The centre frequencies are derived from the subbanded search pipeline. The purple-shaded region shows the on-source integration time. We show the Poissonian uncertainties in the burst rate for individual epochs.}\label{fig:subband_burstrate}
\end{figure*}

We first observed the source with the UWL receiver $\sim$6 days after an Astronomer's Telegram from the CHIME/FRB collaboration announcing the discovery of the source and its repetitions\citep{Shin:2024, Shin:2025}. The first epoch of observation with the UWL resulted in the detection of $4$ bursts within a span of $0.98$\,hr on-source time. The activity observed during the first epoch motivated further follow-up of the source. Subsequent observations over the following $\sim$16 months resulted in the detection of \totbursts bursts from an on-source time of \tottime hours, corresponding to an average burst rate of \burstrate \bursth. Extended Data Table \protect\ref{tab:epoch_list} lists the individual observation epochs, burst rates, number of detected bursts, average RM values, average DM values, and number of bursts detected with S/N $>$20. The observations from PX127 (between February 2024 and September 2024) were performed using unallocated time (Green Time; GT) on the telescope. Hence, a uniform coverage of the burst activity was not possible. The observations from the start of October 2024 (MJD 60949) were conducted using a standard proposal, which is why we have denser and more regular observations of the source, enabling improved studies of burst activity and spectro-temporal variability. We show the full-band and subband burst rates across epochs in Panel A of Figure \ref{fig:burst_activity} and Extended Data Figure \ref{fig:subband_burstrate}, respectively. 
 
\subsection{Murriyang search methodology and burst extraction}
To search for single pulses in the UWL dataset, we use a \code{heimdall}-based overlapping subband search methodology developed specifically for UWL observations \citep{Pravir:2021}. The pipeline first converts the PSRFITS format files into \code{sigproc} filterbank files containing only total intensity data.  These are searched for bursts using  \code{heimdall} \citep[][]{Barsdell:2012}. We search each epoch for single pulses in the DM range of 100-1200 \DMunits. The 3328 MHz bandwidth of the UWL receiver is divided into 52 overlapping subbands of varying widths from 3328 to 64 MHz to search for single pulses. 
We place S/N thresholds of $7.5$ for each subband for a candidate to be considered in  the first evaluation stage. We search for bursts to a maximum boxcar width of $262$\,ms. The candidates generated from this search methodology are classified through \code{fetch} \citep{Agarwal:2019} to reduce spurious events. The results of this classification are vetted by eye to confirm each burst. We also  manually classified all candidates within the DM range of 520 pc \DMunits and 540 \DMunits to search for low S/N bursts that may have been missed by \code{fetch}. The bursts identified can be duplicated at times due to the subbanded search. Hence, we use the \code{heimdall}-reported timestamps of each burst, which are referred to the highest frequency in the given subband, and classify them as unique bursts if they are separated by $>$1 sec. These were further verified by a human to ensure that there is no duplication of bursts. For our analysis, assuming a canonical pulse width of 1 ms, mean system equivalent flux density (SEFD) of 30 Jy for the UWL \citep{Hobbs:2019}, and 7.5 S/N threshold, we estimate the fluence completeness threshold to be $0.6\,(\Delta \nu_{\rm burst}/64 \mathrm{MHz})^{-0.5}$ Jy\,ms, where $\Delta \nu_{\rm burst}$ is the burst bandwidth.
No bursts were detected at higher frequencies from the MARS receiving system. While observing with the MARS receiver, we used a pulsar, PSR~J1644$-$4559, as a test source to verify receiver performance and test search pipelines.

\begin{figure*} 
        \includegraphics[width=1\columnwidth]{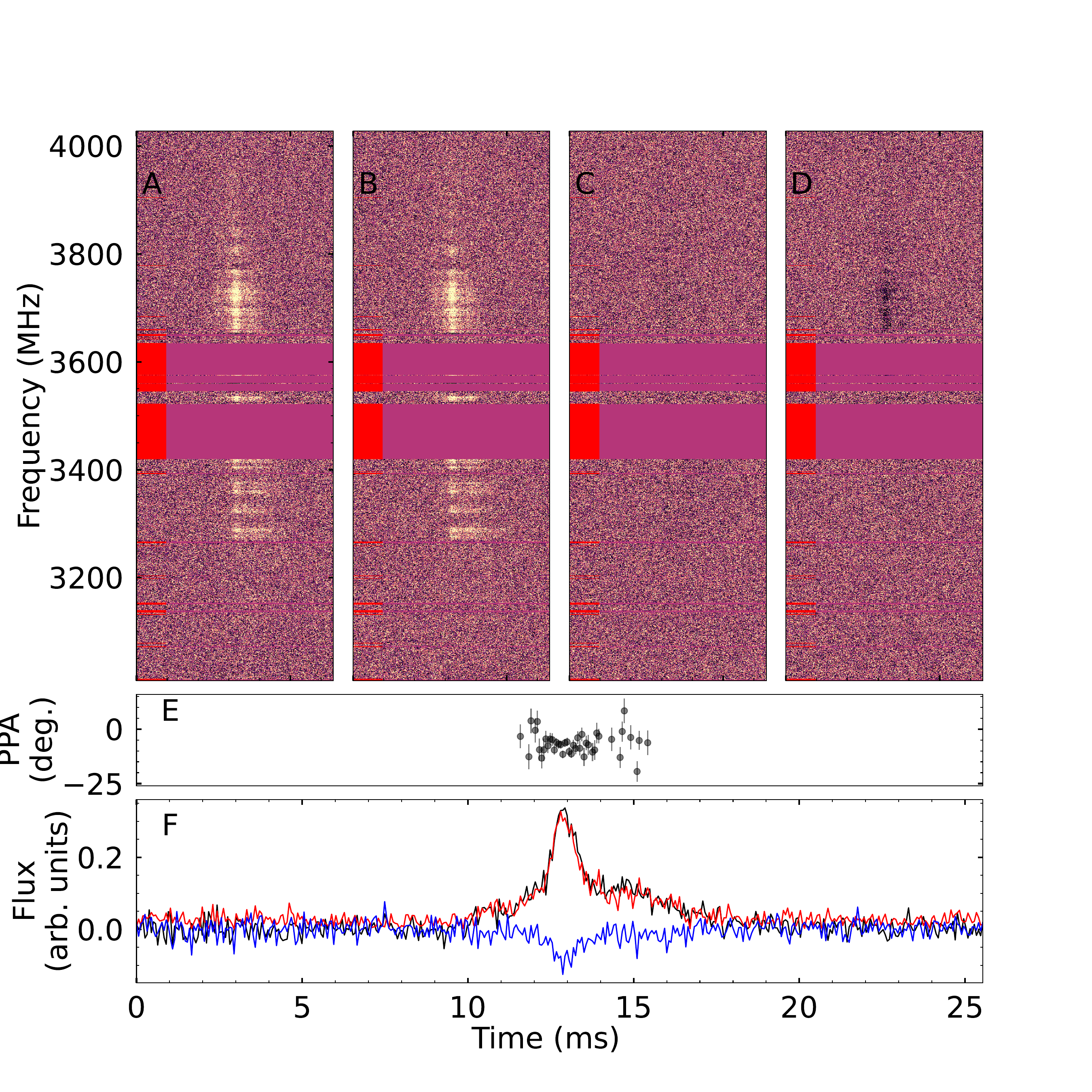}
        \caption{An example FRB20240114A burst from the MJD 60486 epoch. Panels A, B, C, and D show the dynamic spectra for Stokes-I, Q, U, and V, respectively, for the brightest burst from MJD 60486. The dynamic spectra have temporal and spectral resolutions of 64$\mu$sec and 0.5 MHz, respectively. Panels E and F show the PA angle and frequency-averaged pulse profile, respectively. The PA angles are shown for time bins with S/N $>$4. The error bars for PA angles are smaller than the individual markers. The frequency-averaged linear and circular polarisation profiles are shown in red and blue, respectively. Frequency channels that have been excised due to RFI contamination are denoted by a red bar.}
        \label{fig:FRB20240114A_20240626}
\end{figure*}

We use \code{dspsr} \citep{vanStraten:2010} to extract burst archive files in \code{PSRFITS} fold mode format from the raw \code{PSRFITS} search mode data. Individual archive files are manually verified to ensure the quality of the extracted dataset. The extraction of the candidates is done at the candidate arrival times reported by \code{heimdall} \citep{Barsdell:2012}. Inter-channel dispersion is corrected at this stage. Each single pulse archive is RFI flagged using the RFI weights generated by the search pipeline, based on the statistics of the data (see \citep{Kumar:2023} for a detailed description). We calibrate the flux density and polarisation of individual extracted bursts using the \code{psrchive} tool \code{pac} \citep{Hotan:2004}. We use standard flux and polarisation calibration solutions from the Parkes Observatory\footnote{\url{https://www.parkes.atnf.csiro.au/observing/Calibration_and_Data_Processing_Files.html}} to flux calibrate the data and correct for the non-orthogonality of the feed. We extract a  $10.48$\,s window of data around each burst arrival time, with sub-integrations of length $1.048$\,s  each, and manually identify the sub-integration and frequency range of the burst emission from the 3328 MHz bandwidth of the UWL. We show an example extracted burst with polarisation and flux calibration in Extended Data Figure \ref{fig:FRB20240114A_20240626}. Panels A, B, C, and D show the Stokes-I, Q, U, and V dynamic spectra, respectively. The bottom Panel shows the frequency-averaged polarisation profile, which shows a significant circular polarisation. The middle Panel shows the polarisation position angle (PPA).

\subsection{Polarimetry}
\label{sec:polarisation_measurment}
The propagation of radio waves through a cold magnetised medium induces a phase shift in the polarisation position angle (PPA;$\psi$) of the radiation, which can be expressed as 
\begin{equation}
    \psi = {\rm RM} (\lambda^2 - \lambda_o^2),
\end{equation}
where $\lambda$ is the wavelength of the spectral band, $\lambda_o$ is the reference wavelength, and RM is the rotation measure. The rotation measure can be expressed in terms of physical units as 
\begin{equation}
    {\rm RM} = \frac{e^3}{ 2\pi c^4 m_e^2 } \int_{d}^{0} \frac{n_e B_{||}}{(1+z)^2} dl, 
\end{equation}
where $e$ is the charge of the electron, $m_e$ is the mass of the electron, $c$ is the speed of the light, $B_{||}$ is the magnetic field component parallel to the line of sight observer and source, measured as observed seen by the source, $n_e$ is the density of charged particles, and $z$ is the redshift of the charged particles. 

We measure the polarisation properties of the bursts that have a detection S/N$>$20, and for all the bursts for epochs that have a burst rate of $<$10 burst hr$^{-1}$. We show individual RM of the bursts in grey colour points in Panel C of Figure \ref{fig:burst_activity}. We use RM Synthesis \citep{Brentjens:2005} from \code{rm-tools} \citep{Purcell:2020}\footnote{https://github.com/CIRADA-Tools/RM-Tools} to estimate the RM values of each of the polarisation and flux calibrated bursts.

\begin{figure}[h]
    \centering
    \begin{subfigure}[b]{0.9\textwidth}
        \centering
        \includegraphics[width=\textwidth]{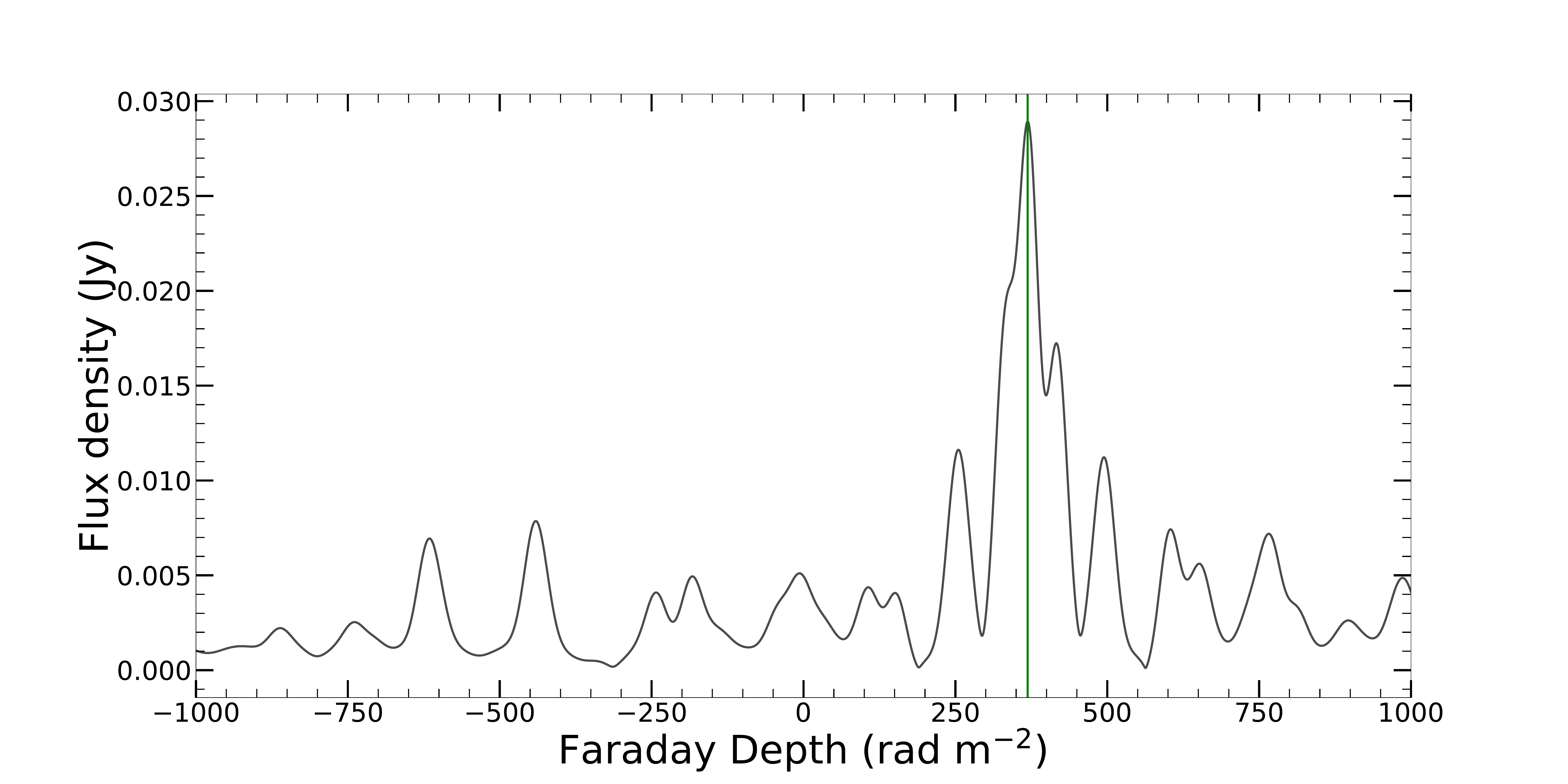}
        \caption{}
        \label{fig:FRB20240114A_Pol_1_sub1}
    \end{subfigure}\\
    \begin{subfigure}[b]{0.9\textwidth}
        \centering
        \includegraphics[width=\textwidth]{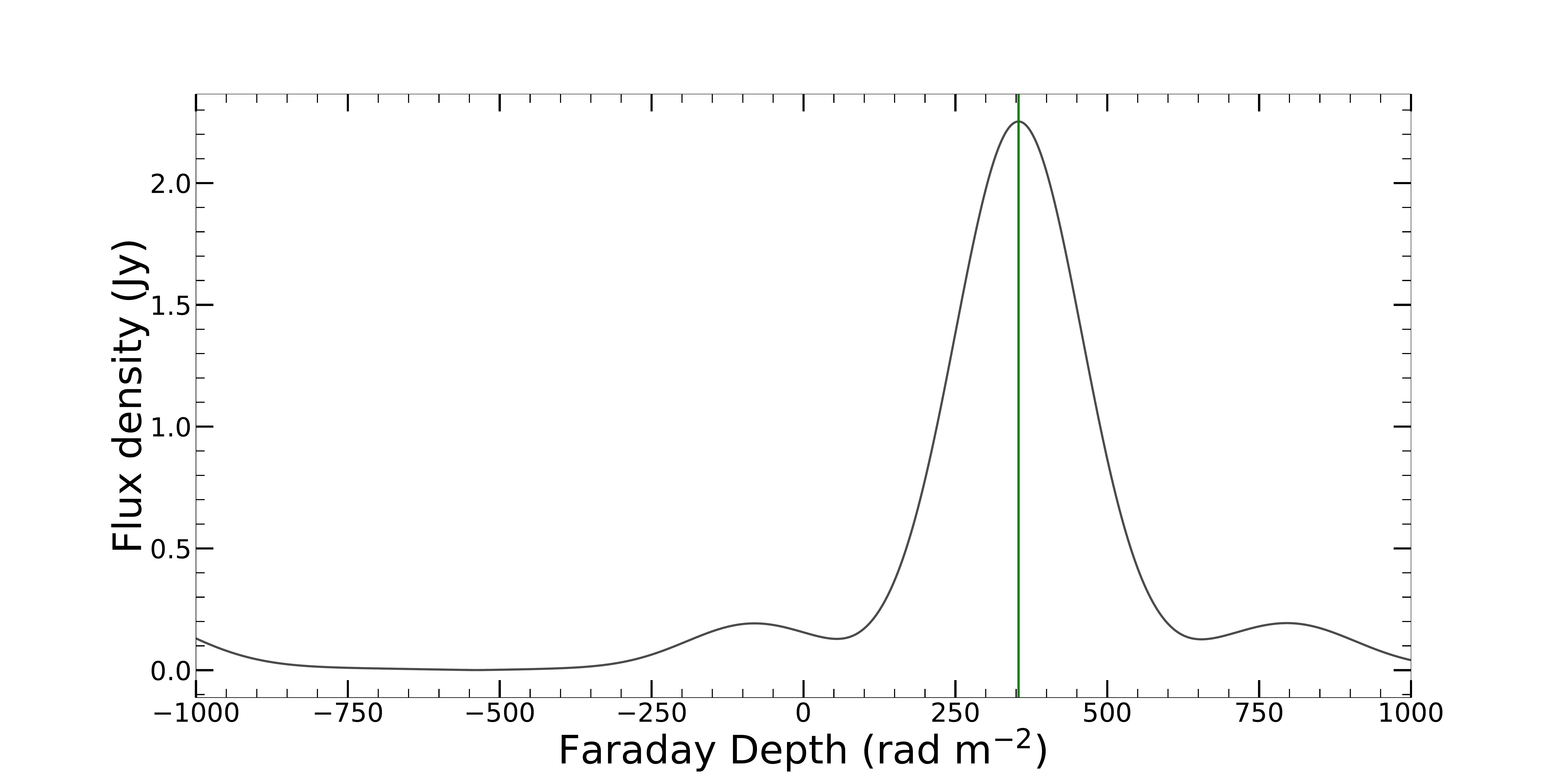}
        \caption{}
        \label{fig:FRB20240114A_Pol_1_sub2}
    \end{subfigure}
    \caption{Faraday depth profiles of bursts from FRB~20240114A.
    We show the FDF profiles between -1000 and +1000 \protect\RMunits. The green colour vertical line shows the peak value of the clean FDF obtained from \protect\code{rmclean}.  The bursts \protect\ref{fig:FRB20240114A_Pol_1}(a) and \protect\ref{fig:FRB20240114A_Pol_1}(b) were detected at MJDs 60645 and 60420, respectively.}\label{fig:FRB20240114A_Pol_1}
\end{figure}

To estimate the RM, we first frequency-average the profile using a boxcar profile with a 50\% pulse width. 
We first create Stokes Q and U profiles using the extracted data from the archive files for each burst. We smooth the profiles using a Gaussian filter with  spectral and temporal widths of 1 MHz and 4 ms, respectively. We average the Gaussian-smoothed profiles between 50\% of the frequency-averaged peak value from the Stokes-I profile. This 1-D Stokes profile is then used to run \code{rmsynthesis} and \code{rmclean}. We show the measured RMs from \code{rmsynthesis} for bursts with S/N $>$20, for each epoch in Panel C of Figure \ref{fig:burst_activity}. The black points show the RM of all the bursts with S/N$>$20, and the red points show the average RM values of specific epochs. Example Faraday Dispersion Functions (FDFs) are shown in Extended Data Figure \ref{fig:FRB20240114A_Pol_1}.
We estimate the error on the average RM using propagation of uncertainties from individual RM measurements for a given epoch. We measure an average RM of \avgRM \RMunits throughout the campaign. We also observe a large scatter in RM values on some of the epochs (e.g., a standard deviation of 10.7 \RMunits on MJD 60685). We note that RMs for some bursts that show significant deviation from the mean RM are generally associated with high circular polarisation. 

Higher levels of circular polarisation can be due to Faraday conversion, which can lead to a biased estimate of the RM value \cite[e.g.,][]{Uttarkar:2024}. However, any such effect will be averaged for epochs that have a large number of bursts, as only a small fraction ($<$10\%) in individual epochs show significant circular polarisation. Such transient circular polarisation has been observed in other FRBs such as FRBs~20180301A and 20201124A \citep[e.g.,][]{Price:2019, Uttarkar:2023, Jiang:2024}.
To search for secular variation in RM,  we fit for linear drifts in RM separately for epochs between MJDs 60342 - 60652 and  60653 - 60827. 
We observe a modest long-term decreasing trend in the RM of the source over our follow-up period.
The epochs until MJD $\sim$ 60654 showed a small increase in the measure of RM of 0.17$\pm$0.001 \RMunits day$^{-1}$. However, epochs after MJD $\sim$60654 show a more pronounced decrease of 0.9$\pm$0.001 \RMunits day$^{-1}$. The straight-line fit to the evolution of RM is shown in Panel C in Figure \ref{fig:burst_activity}. 

We use the estimated RM to de-rotate the spectra to account for Faraday rotation \cite[e.g.,][]{Uttarkar:2024}
\begin{equation}
    \begin{split}
        Q_{\rm de-rot}  =   Q \cos(2\psi)  + U \sin(2\psi),\\
        U_{\rm de-rot}  =   U \cos(2\psi)- Q \sin(2\psi),
    \end{split}
\end{equation}
where $Q_{\rm de-rot}$ and $Q_{\rm de-rot}$ are the Faraday de-rotated spectra. We frequency average Stokes-Q and  Stokes-U de-rotated spectra to calculate PPA and ellipticity angle (EA; $\chi$) using
\begin{equation}
    \psi = \frac{1}{2} {\rm tan}^{-1}\left(\frac{U_{\rm favg}}{Q_{\rm favg}}\right)
    \label{eqn:PA_cal}
\end{equation}
and
\begin{equation}
    \chi = \frac{1}{2} {\rm tan}^{-1}\left(\frac{V_{\rm favg}}{\sqrt{Q^2_{\rm favg}+U^2_{\rm favg}}}\right). 
    \label{eqn:EA_cal}
\end{equation}

We measure the errors associated with $\psi$ and $\chi$ \citep[e.g.,][]{Day:2020, Uttarkar:2023}  using
\begin{equation}
    \sigma_{\psi} = \sqrt{\frac{Q_{\rm favg}^{2}\sigma_Q^2+U_{\rm favg}^{2}\sigma_U^2}{4(Q_{\rm favg}^2+U_{\rm favg}^2)^2}}
    \label{eqn:PA_err_cal}
\end{equation}
and
\begin{equation}
    \sigma_{\chi} = \frac{\sqrt{(\Vf\Qf)^2\sigma_Q^2 + (\Vf\Uf)^2\sigma_U^2 + (\Qf^2+\Uf^2)^2\sigma_V^2}}{2(\Qf^2+\Uf^2+\Vf^2)\sqrt{\Qf^2+\Uf^2}},
    \label{eqn:EA_err}
\end{equation}
where $V_{\rm favg}$, $U_{\rm favg}$, and $Q_{\rm favg}$ are the frequency-averaged de-rotated spectra, and $\sigma_U$, $\sigma_{Q}$, and $\sigma_{V}$ are the uncertainties associated with the Stokes U, Q, and V spectra, respectively.

\subsection{Dispersion measure estimation}
\label{sec:DM_measurments}
The estimation of DM in FRBs is particularly nuanced due to the complex spectral structure seen in repeating FRB sources (e.g., the sad trombone structure and band-limited emission \citep[][]{Hessels:2019, Kumar:2023}). The S/N maximisation of the burst alone can lead to a biased estimate of the DM, since it will mask the underlying structure of the burst. Hence, we rely on the structure maximisation method to estimate DM. However, we report DMs measured by both S/N maximisation and structure maximisation methods, for completeness (see Extended Data Table \ref{tab:epoch_list}). 
The DM measurement of the bursts is performed using \code{DMPhase} and \code{pdmp}. \code{DMPhase} uses coherent power to estimate the DM, such that it maximises structure, rather than the S/N of the burst. \code{DMPhase} has been widely used in FRB DM measurements  \citep[e.g.,][]{Kumar:2023}. In addition to the \code{DMPhase}, we also use match-filtered boxcar estimation of the S/N using \code{pdmp}. We use the extracted archives, which have been cropped to include only the time and frequency bandwidth where burst emission is present. The extracted archives are integrated to a time resolution of 0.25 ms. We use a Gaussian filter with a temporal width of $0.25$\,ms and a spectral width of $2$\,MHz to smooth the profile before we estimate the burst DM. We estimate this for bursts that have an S/N $>20$. We use visual inspection to ensure that the archives are free of any RFI channels. However, despite a concentrated effort to remove all RFI-affected channels, some archives remain with residual RFI. The residual RFI  is particularly an issue with large-bandwidth instruments such as the UWL. Hence, to avoid biasing our measurement of the DM, we reject 45 bursts due to residual and broadband RFI out of the total of  1226 bursts  with S/N$>20$. The averaged DM for a given epoch measured by \code{DMPhase} and \code{pdmp} are listed in Extended Data Table \ref{tab:epoch_list}. The uncertainties reported for each epoch are estimated using error propagation of uncertainties. The DM values show a wide distribution of measured values (as shown in Panel D of Figure \ref{fig:burst_activity}). We also observe short-term variations in the DM values during our follow-up campaign. We consider two burst storm periods, B4 and B5, between MJDs 60610 to 60725 and 60750 to 60825, respectively, to estimate the excess DM. During the burst storms of interest, we estimate an average structure maximised DM of 530.37$\pm$0.34 \DMunits and 529.62$\pm$0.35 \DMunits, for B4 and B5, respectively. We measure the standard deviation of the DM values to be 4.75 \DMunits and 5.22 \DMunits for B4 and B5, respectively. We estimate an average DM of \avgDM \DMunits with a standard deviation of 9 \DMunits for the whole campaign using \code{DMPhase}. We report the structure maximised DM for the bursts above an S/N of 20\footnote{\url{https://github.com/pavanuttarkar/FRB20240114A/blob/main/FRB20240114A_20sig_properties.pdf}}.

\subsection{Fluence and pulse width measurements}
\label{sec:fluence_width_measurments}

A Bayesian approach is used to measure burst fluences and widths.
The on-pulse region of the pulse is estimated using a boxcar: 
\begin{equation}
    \mathcal{T} =  \begin{cases}
        1,\,&\text{if } \,t_0<t<t_1 \\
        0,              & \text{otherwise},
    \end{cases}
    \label{eqn:boxcar_eqn}
\end{equation}
where 
$t_0$ and $t_1$ are the start and end time, respectively.

We first subtract the baseline in the flux and polarisation calibrated archives using the \code{remove\_baseline} method in \code{psrchive}. The bursts are then frequency-averaged using \code{fscrunch} utility in \code{psrchive}. 
The pulse is then fitted with a boxcar defined by Equation \ref{eqn:boxcar_eqn}. The fluence of the burst is then calculated as
\begin{equation}
    \mathcal{F} = \sum_{t_0}^{t_1} P_{\rm burst} (t)  \Delta t. 
\end{equation}
where $P_{\rm burst}$ is the frequency-averaged total polarisation profile. 

We use \code{bilby} with \code{dynesty} nested sampler \cite[][]{Ashton:2019} to fit a boxcar to estimate the pulse width.
We assume the noise in the data to be Gaussian and use a  Gaussian likelihood function of the form 
\begin{equation}
\begin{split}
     \textit{L}(P_{{\rm burst}, i}| P_{\rm burst, model} , \sigma) =  \prod_i^{N_f} \left[\frac{1}{\sqrt{2\pi\sigma^2}}\,\,\exp\left(-\frac{(P_{{\rm burst}, i}-P_{\rm burst, model})^2}{2\sigma^2}\right)\right],     
    \label{eq:posterior_likelihood}
\end{split}
\end{equation}
where 
$P_{\rm burst, model}$ is the modelled boxcar and $\sigma$ is the root mean square noise level. 

The inferred fluence-pulse width distribution is shown in Extended Data Figure \ref{fig:FRB20240114A_FLUENCE}. The $>$20 S/N bursts from FRB 20240114A are shown in purple points, samples of bursts from FRBs 20121102A \citep{Hewitt:2022} and 20190520B \citep{Niu:2022} are shown in green and orange points, respectively. The  FRBs detected by ASKAP in the fly's eye survey \citep{Shannon:2018} are shown in grey. We show the fluence and boxcar pulse width distributions in purple. Log-Gaussian fits to the distributions are shown in solid purple lines. The distributions are likely  affected by the selection effects, given the high S/N cutoff for this plot.

\subsection{Physical model for burst spectral variations: Extreme scattering events}

\label{sec:Physical_model}

Our preferred explanation for the spectral and temporal variations observed in the bursts from FRB~20240114A is lensing due to foreground plasma. This lensing is commonly referred to as an extreme scattering event (ESE).
We model this using an approach that has been previously used for quasars \citep{Bannister:2016} and pulsars \citep{Kerr:2018}. In the studies for quasars and pulsars, the lenses were assumed to be in our Galaxy \citep{Bannister:2016, Kerr:2018}.  

We model the ESEs using the Gaussian lens model \cite[][]{Clegg:1998}, which has been extended to FRBs \citep[][]{Cordes:2017}. 
In previous approaches, it was possible to directly relate lens magnification to source brightness.  This is because quasars and many pulsars (when averaged over many pulses) have constant flux density.
In our case, this is not possible.  Instead we model the effect of magnification on the burst rate. When there is higher (lower) magnification, quantified by gain, a higher (lower) burst rate is expected. 

We assume a point source for the ESE modelling, which has implications in terms of the expected ESE light curve (see \cite{Clegg:1998} for detailed discussion). A dimensionless parameter $\alpha$ can model the plasma lens geometry \citep{Cordes:2017}:
\begin{equation}
    \alpha = \frac{\lambda^2 r_e {{\rm DM}_l}}{\pi a^2}\left(\frac{d_{sl}d_{lo}}{d_{so}}\right) = 3430 \left(\frac{{\rm DM}_l}{\rm pc\,cm^{-3}}\right)\left(\frac{d_{\rm sl}}{\rm kpc}\right)\left(\frac{\nu^{-2}}{\rm GHz}\right)\left(\frac{a^{-2}}{\rm au}\right),
    \label{eq:len_alpha}
\end{equation}
where $\lambda$ is the wavelength, $a$ is the lens scale, ${{\rm DM}_{l}}$ is the dispersion measure depth, and $d_{so}$, $d_{sl}$, and $d_{lo}$ are distances from the source to the observer, source to the lens, and lens to the observer. When the lens is in the host galaxy, the parameter $\alpha$ is dominated by $d_{sl}$ as $d_{lo}/d_{so}\sim1$.

The gain induced by the lens is 
\begin{equation}
    G = |1 + \alpha(1-2u^2)\exp{(-u^2)}|^{-1},
    \label{eqn:Gain}
\end{equation}
where $u$ is the transverse scale in the lens frame (see \citep{Cordes:2017} for a detailed explanation).

The gain of the system is estimated by solving for the roots of 
\begin{equation}    
    u(1+\alpha e^{-u^2})=u^\prime,     
    \label{eq:roots}
\end{equation}
where $u^\prime$ is the transverse offset in the observer reference frame. 

We can then connect the gain to the  burst rate through the burst S/N distribution. For this, we define S/N as $\mathcal{S}$.
We assume that the intrinsic form of the S/N distribution does not change between epochs.
We find that the burst S/N distribution is well modelled at all epochs by a power law $dN/d\mathcal{S}  \propto \mathcal{S}^{-\gamma}$, where $N$ is the number of bursts and $\mathcal{S}$ is the S/N of the bursts.
We fitted a power law function to the S/N differential distribution for individual epochs and estimated the power law index $\gamma$ to be 2.8. 
The burst rate will then vary as the source is magnified or de-magnified (characterised by a gain parameter $G$), which affects the number of bursts above an S/N threshold.  These two quantities are related using
\begin{equation}
    R = A\int_{S_{\rm min/G}}^{\infty} {\mathcal{S}}^{-\gamma} d{\mathcal{S}},
    \label{eqn:lum}
\end{equation}
where $S_{\rm min}$ is the baseline S/N of the bursts, A is the scaling factor, 
and $G$ is the gain factor that changes the S/N of a burst. 

We can estimate the gain by  Equation \ref{eqn:lum}:
\begin{equation}
          R=\frac{A}{1-\gamma}\left(\frac{S_{\rm min}}{G}\right)^{-\gamma+1}.
\end{equation}

We can then solve this for the gain:
\begin{equation}
        G \propto R^{1/(\gamma-1)}.
\end{equation}
The uncertainty on the estimated gain, calculated using error propagation, is 
\begin{equation}
    \sigma_G = \frac{R^{1/(\gamma-1)}}{(\gamma-1)\sqrt{N}},
\end{equation}
where $N$ is the number of bursts.\\

\begin{figure}[h]
    \centering
    \begin{subfigure}[b]{1\textwidth}
        \centering
        \includegraphics[width=\textwidth]{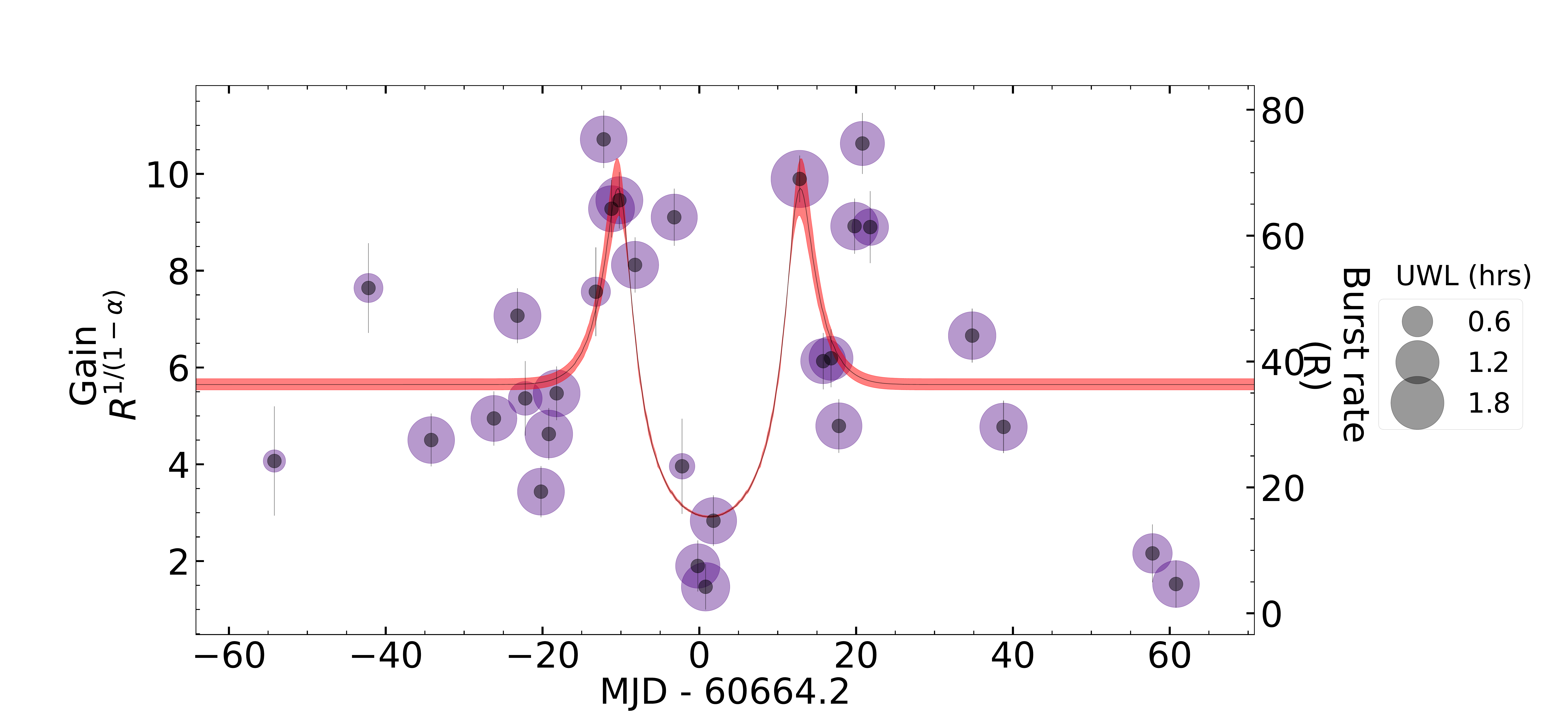}
        \caption{}
        \label{fig:burst_storm_individual_sub1}
    \end{subfigure}\\
    \begin{subfigure}[b]{1\textwidth}
        \centering
        \includegraphics[width=\textwidth]{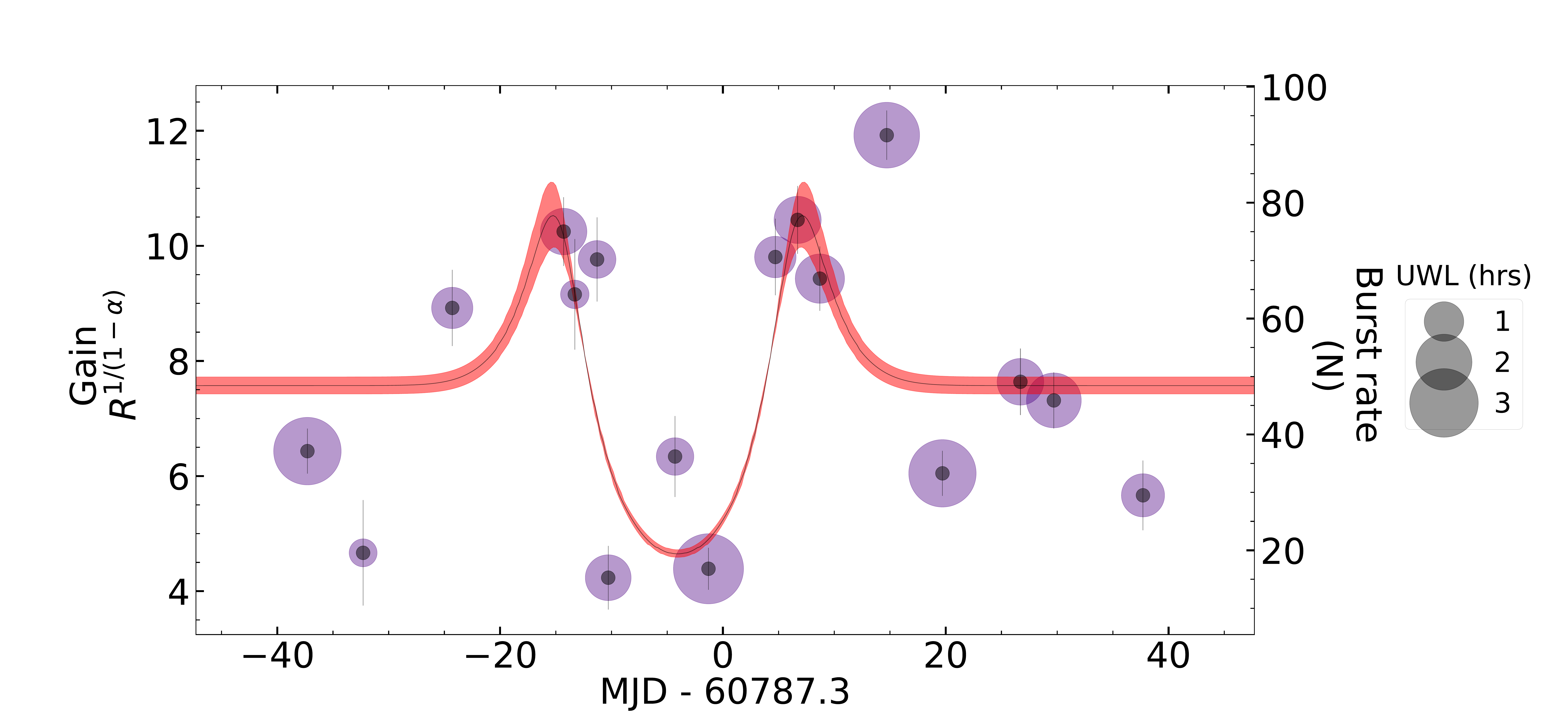}
        \caption{}
        \label{fig:burst_storm_individual_sub2}
    \end{subfigure}
    \caption{ESE burst rate modelling. We show the light curve fit to the gain function derived from the burst storms for burst storms B4 and B5 in Extended Data Figures \ref{fig:burst_storm_individual}(a) and (b), respectively. The 1-$\sigma$ error region for the curve fit is shown in red. The on-source time is shown in shaded purple for each epoch.  The fit parameters are listed in Extended Data Table \ref{tab:len_params}. The burst rate is shown as black points, and the errorbars are the 1-$\sigma$ uncertainties of the gain function for each epoch.}
    \label{fig:burst_storm_individual}
\end{figure}

We focus our modelling on observations after MJD 60610, where we have more uniform coverage. 
We consider the two burst storms in the MJD intervals 60610-60728 and 60750-60827. Since finding the roots of a non-linear equation like Equation \ref{eq:roots} is non-trivial and extremely time-complex, we use a \code{cuda}-based graphics processing unit (GPU)-accelerated fit routine to fit the model, using \code{cupy} \citep{cupy_learningsys:2017}. We use a Bayesian method for the model fit using \code{bilby} and \code{DYNESTY} nested sampler. 
A Gaussian likelihood, similar to that presented in Equation \ref{eq:posterior_likelihood}, is used to model the gain. The posterior distributions for the lens parameters are shown in Extended Data Figure \ref{fig:posteriors_ESE}. The physical interpretation of the burst parameters is discussed in the main text. We show the individual lens fit for burst storms B4 and B5 in Figure \ref{fig:burst_storm_individual}. The lens fit parameters are listed in Extended Data Table \ref{tab:len_params}.

A schematic model for such a scenario is shown in Figure \ref{fig:ESE_toy}. Panel A shows the density variation in the plasma, which refracts the incident plane waves. The circumsource medium is likely to be chaotic. This is shown in the dashed lines in Panel A. Although we model the two burst storms as pure Gaussian lenses, the plasma overdensities and underdensities are expected to be turbulent. Some of the stochastic variation in the burst rate can be attributed to such a scenario. Hence, we attribute the deviation of the burst rate from the model to such turbulence. 
The ray-trace plot for the waves refracted by the Gaussian lens is shown in Panel B. The integrated lensing effect, which leads to an apparent change in the brightness of the source is shown in Panel C. 


\subsection{Searching for periodic activity}
\label{sec:periodicity}

FRBs such as 20121102A and 20180916B have been observed to exhibit periodic activity windows of $\sim$160 days and $\sim$16 days, respectively \citep{Rajwade:2020, Chime/FRBCollaboration:2020}. Such periodic activity windows have motivated possible progenitor models \citep[e.g., a Be/X-ray binary system;][]{Li:2021}. Other FRBs have also been claimed to show possible periodic behaviour \citep[e.g., FRB 20240209A;][]{Pal:2025}. However, 
testing for periodic behaviour requires a large sample set of bursts and a dense temporal follow-up of the source. Our observations with Murriyang fulfil both criteria, making it an ideal candidate to search for any periodicity. This is further motivated by the regular variations in burst activity with time, every $\sim 70$\,days. We searched for long-timescale periodicity using the Lomb-Scargle \citep{Lomb:1976, Scargle:1982} method and fast folding algorithm (FFA).

The Lomb-Scargle periodogram was calculated using
\begin{equation}
    \begin{split}        
    P_{\rm LS}  =  \frac{1}{2}\left(\sum_{N} B_n \cos{2\pi f (t_n - \tau)}\right)^2/\sum_{N}\cos^2{2\pi f (t_n - \tau)}\\
        + \frac{1}{2}\left(\sum_{N} B_n \sin{2\pi f (t_n - \tau)}\right)^2/\sum_{N}\sin^2{2\pi f (t_n - \tau)},
    \end{split}
\end{equation}
where $B_n$ is the burst rate for each epoch, $\tau$ is a  function of $f$ and the observing epochs $t_n$: 
\begin{equation}
    \tau = \frac{1}{4\pi f} \arctan\left(\frac{\sum_N \sin{(4\pi f t_n)}}{\sum_N \cos{(4\pi f t_n)}}\right).    
\end{equation}

\begin{figure}
    \flushleft
    \includegraphics[width=1.1\linewidth]{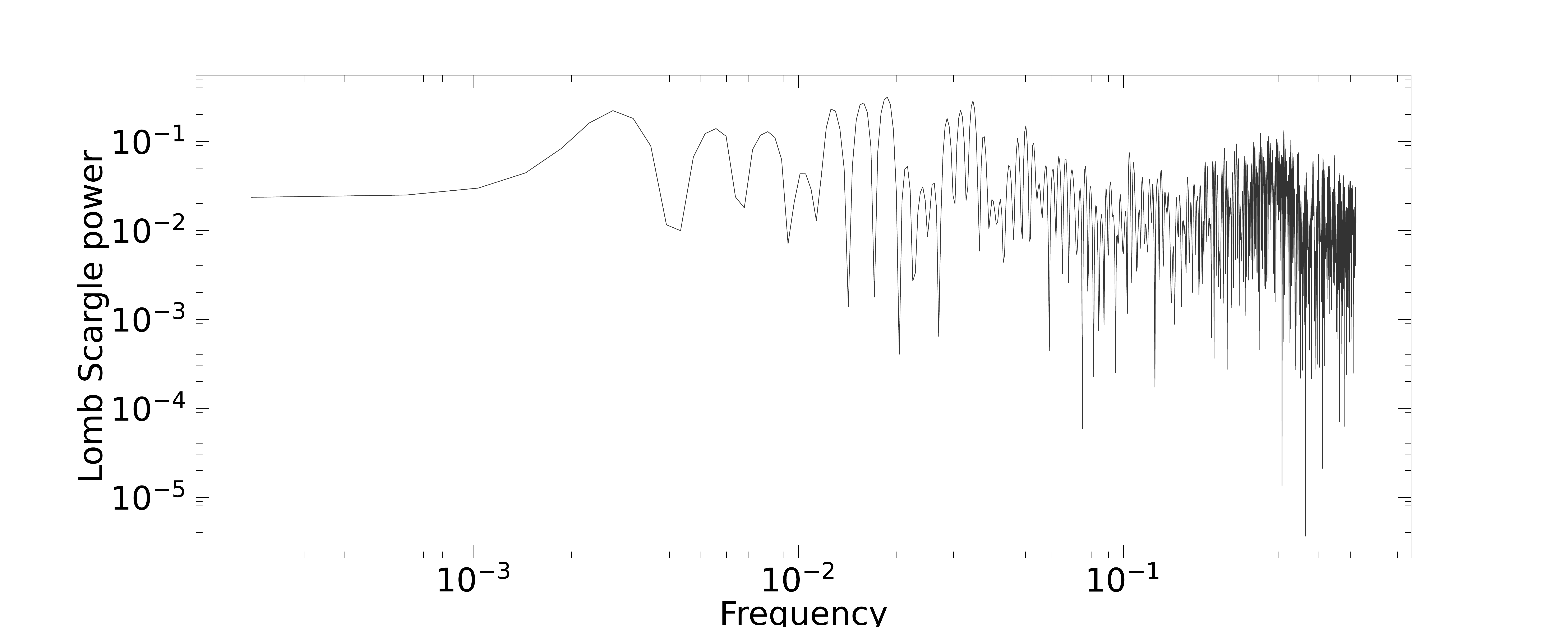}
    \caption{Lomb-Scargle periodogram of burst rate. The unequally sampled observational data were searched for periodicity with the \code{astropy} Lomb-Scargle method\protect\footnotemark \protect\citep{Lomb:1976}. The plot is shown in log-log scale space. No significant evidence for periodicity is recovered in the Lomb-Scargle search. }
    \label{fig:LS}
\end{figure}

We identify a periodicity of $\sim$53 days with a p-value of 0.02 using the Lomb-Scargle method based on the burst rates from our observing campaign. 
The Lomb-Scargle periodogram is shown in Extended Data Figure \ref{fig:LS}. 

For the FFA search, we follow the approach previously used to search for periodicity in FRB~20121102A \cite[][]{Rajwade:2020}. We use burst arrival times as reported by the detection pipeline to bin them into $150$ equal bins, each of 74 hours. We then search for periodicity using the FFA search. 
We attempted different bin widths to search for periodicity. However, we do not find any significant evidence for periodicity in the FFA when we consider all the epochs together above an S/N of 8. 

We attribute the significant detection of periodicity in the Lomb-Scargle search to the quasi-periodic nature of variable activity, which explains the non-detection of a period in the FFA search.
It is possible that a significant periodicity would be identified in a longer data set.

\subsection{Spectro-temporal and polarisation properties of FRB~20240114A}
\subsubsection{Spectral morphology: Spectral memory}
FRB~20240114A shows spectral variations on a range of time scales. Figure \ref{fig:ESE_evolution} shows spectral variations on day-week time scales. The source also exhibits spectral memory in its burst emission, with timescales ranging from milliseconds to minutes.  We refer to these bursts as ``carbon copies''. They often show a replica of the preceding burst, separated by $\sim$tens of milliseconds to $\sim$tens of minutes. 
We show the millisecond-duration separated bursts in Extended Data Figure \ref{fig:ms_carbon_copies}. These carbon copies have not been observed to correlate with any phases of the burst activity. The long-timescale carbon copies are shown in Extended Data Figures \ref{fig:narrow_carbon_copies} and \ref{fig:narrow_carbon_copies1}. The central frequency of the bursts  generally varies amongst observing epochs. We list the spectral bandwidth, central frequency, and the burst epochs in Extended Data Table \ref{tab:spec_mem}. 
We detect minute-scale carbon copies in multiple epochs, notably at a different frequency. Narrow band burst emission has previously been interpreted as a plausible indication of plasma lensing \citep{Hessels:2019,Gourdji:2019,Pravir:2021}. 

\subsubsection{Polarisation properties: Polarisation Position Angle}

Burst polarimetry, in particular burst PPA, can be used to understand the origin of FRB emission better.
We show examples of bursts that show sub-burst  PPA variations in Extended Data Figure \ref{fig:PA_swing}. We show the long-term PPA and EA evolution of the source in Extended Data Figures \ref{fig:PA_evolution} and \ref{fig:EA_evolution}, respectively.  We consider only bursts with S/N $>$ 20 in Extended Data Figures \ref{fig:PA_evolution} and \ref{fig:EA_evolution}. 
The individual EA and PPA points have an S/N$>$4. The long-term monitoring of the PPA shows a multi-peak distribution. We fit a multi-peak Gaussian distribution to the histogram using \code{bilby} and \code{DYNESTY} sampler. We estimate the peaks of the distribution at -17.39$^{+0.5}_{-0.5}$ degree, $<$21.7 degree, and 64.8$^{+0.6}_{-0.6}$ degree. Panels A of Extended Data  Figures \ref{fig:PA_evolution} and \ref{fig:EA_evolution} show PPA and EA for individual time bins with a time resolution of \PAtimebin. Panel B shows the distribution of the PPA and EA with a resolution of 0.8$^\circ$.  We show the multi-component Gaussian function fit to the data in Panel B of Extended Data Figure \ref{fig:PA_evolution} in black. The individual PPA points are colour-marked with the centre frequency of the bursts.

Most repeating FRBs generally show a high degree of linear polarisation and a flat PPA angle across the on-pulse region \cite[e.g.,][]{Day:2020, Kumar:2023}. Only a small fraction of the bursts from both repeating and non-repeating sources have been seen to have a sweep in the PPA angle or the S-curve, similar to the behaviour seen in pulsars \cite[][]{Luo:2020,Mckinven:2024}. The observed S-curve in pulsars can be attributed to the rotation of a neutron star, which projects the magnetic field axis across the line-of-sight using the rotating vector model \cite[RVM, Ref][]{Rashakrishnan:1969}. Similar morphological characteristics in the PPA angle for FRBs were argued to be due to an RVM-like scenario \cite[e.g.,][]{Luo:2020, Mckinven:2024} because of the rotation of a neutron star. Such behaviour was used to argue for a magnetospheric origin of FRBs \citep{Luo:2020}. However, studies of repeaters and non-repeaters suggest that the scenario is more complex, with PPA variation likely driven by a dynamically evolving magnetosphere \citep{Xiaohui:2025}. 

FRB~20240114A is consistent with the wider FRB population in that a significant fraction of the bursts show a flat PPA angle, while only a handful of the bursts show variation in the PPA angle.


\begin{figure*} 
        \centering
        \includegraphics[width=1\columnwidth]{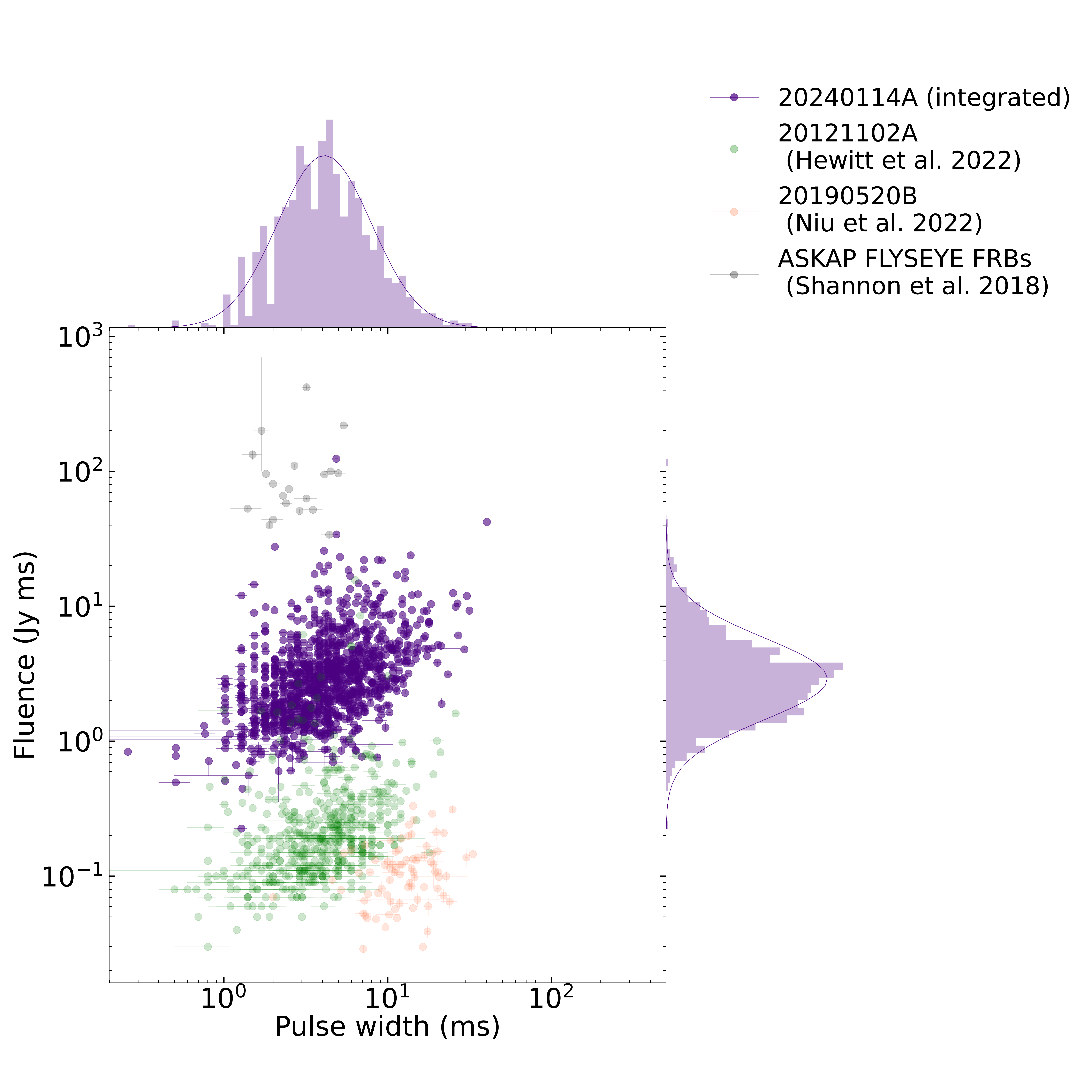}
        \caption{The fluence pulse width relationship. The fluence estimate of the bursts from this study above an S/N of 20 and the estimated pulse width is shown for 20240114A in purple. We also show the pulse width fluence distribution for ASKAP FLYSEYE FRBs \protect\cite[grey;][]{Shannon:2018}, as well as for repeating FRBs 20121102A \protect\cite[green;][]{Hewitt:2022} and 20190520B \protect\cite[red;][]{Niu:2022}. The pulse width for the bursts is estimated using a Bayesian fit to the frequency-averaged Stokes-I profile with a boxcar (see Section \protect\ref{sec:fluence_width_measurments}). The histograms are shown in purple, and the purple line denotes the Log-Gaussian fit to the distribution. }
        \label{fig:FRB20240114A_FLUENCE}
\end{figure*}

\begin{figure}
    \centering
    \begin{subfigure}[b]{0.9\textwidth}
        \centering
        \includegraphics[width=\textwidth]{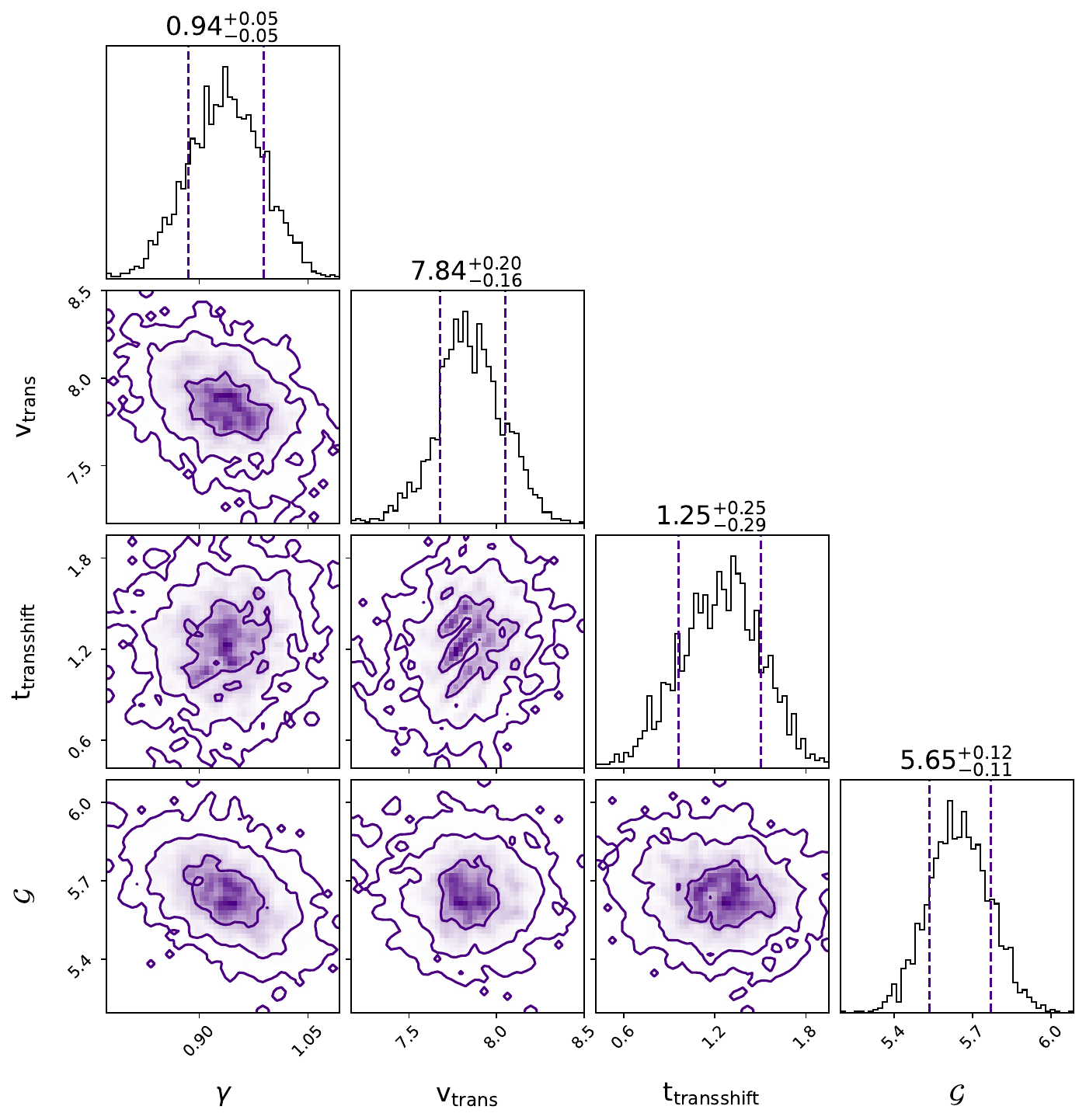}
        \caption{}
        \label{fig:posteriors_ESE_sub1}
    \end{subfigure}

\end{figure}

\begin{figure}
    \ContinuedFloat
    \begin{subfigure}[b]{0.9\textwidth}
        \centering
        \includegraphics[width=\textwidth]{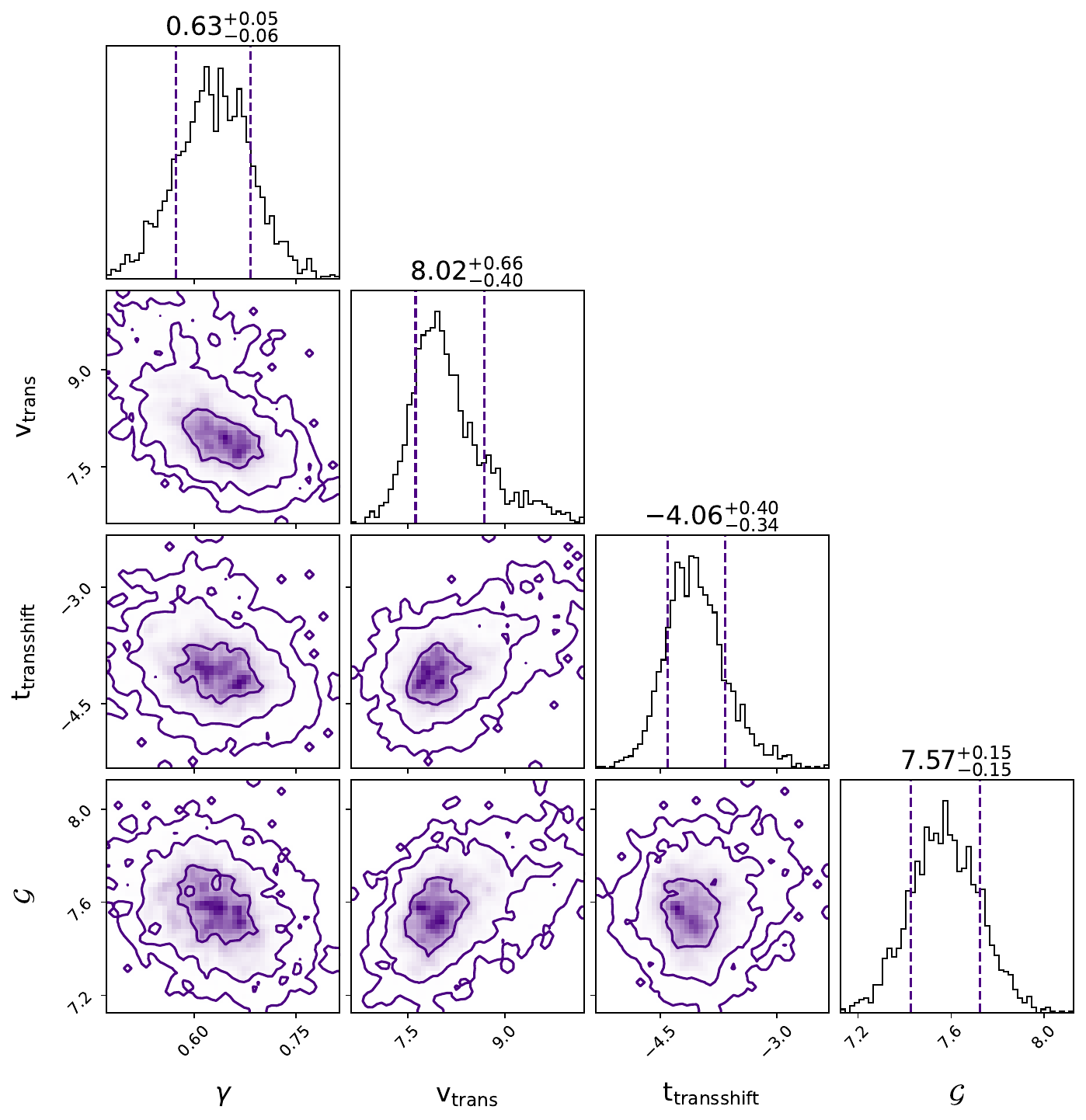}
        \caption{}
        \label{fig:posteriors_ESE_sub2}
    \end{subfigure}
    \caption{Posterior distribution for ESE light curve model. Figures \protect\ref{fig:posteriors_ESE}(a) and (b) show the posteriors for two burst storms, B4 and B5.  The measurements from the light curve fit are reported in Extended Data Table \ref{tab:len_params}. \vspace{2cm}}\label{fig:posteriors_ESE}
\end{figure}

\footnotetext{https://docs.astropy.org/en/latest/timeseries/lombscargle.html}

\setcounter{figure}{8}
\begin{figure*}[!htbp]
    \centering
    \includegraphics[width=1\columnwidth]{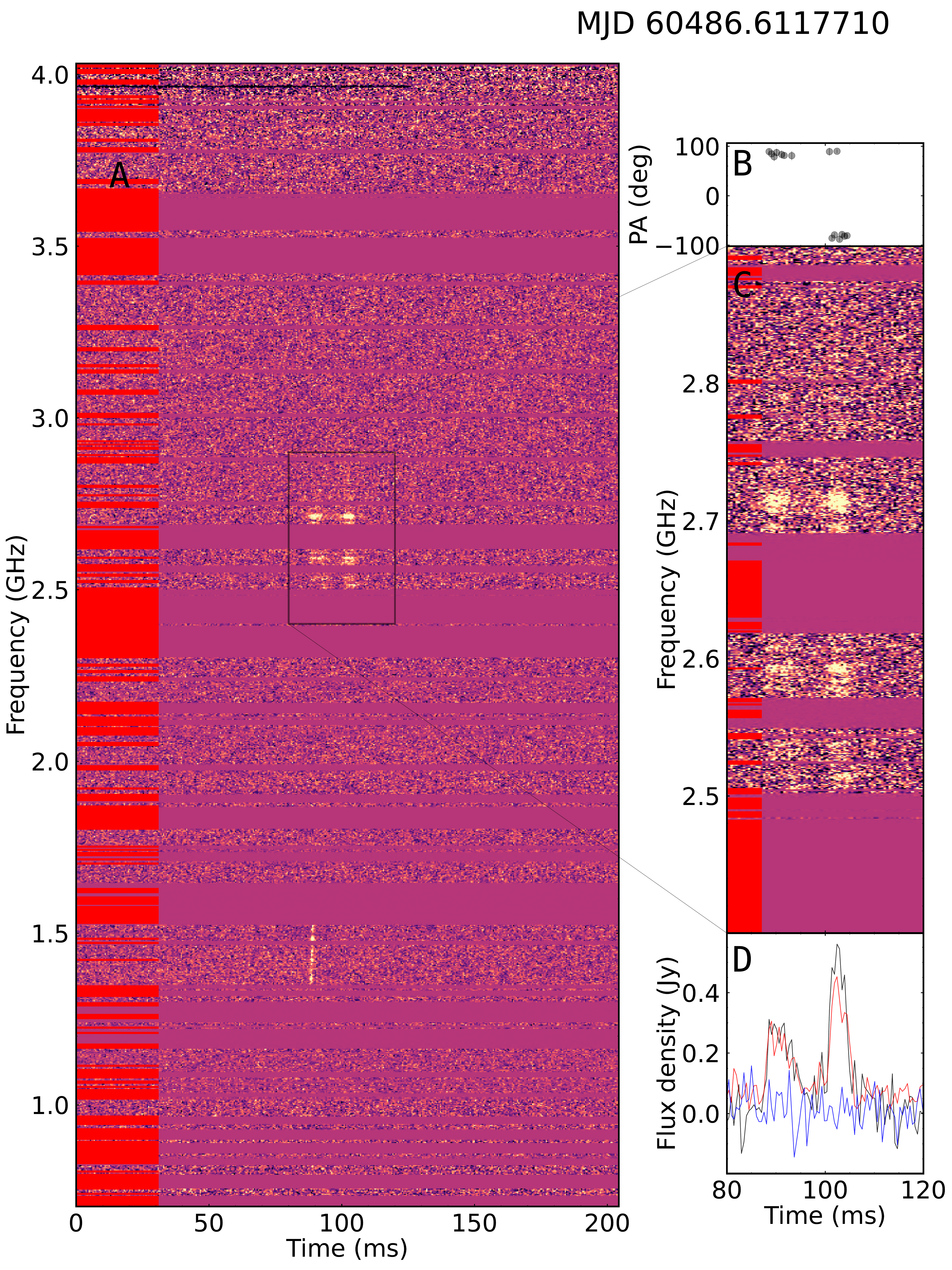}
    \label{fig:ms_carbon_copies}
\end{figure*}

\begin{figure*}[!htbp]
    \ContinuedFloat
    \caption{Short timescale carbon copies of bursts. The bursts, detected on MJD 60750, centred at $\sim$2.68 GHz shows a similar morphology. The burst in the lower part of the spectrum is not spectrally connected to the carbon copied bursts centred at $\sim$2.68 GHz. The subbanded search pipeline independently detected all  bursts. We show the carbon copy bursts in the zoomed-in inset plot. The spectrum for the complete band and the inset plot have temporal and spectral resolution of 0.512 ms and 0.5 MHz, respectively. Panels A, B, C, and D show the dynamic spectrum of the burst with the whole band, the PPA angle, the zoomed-in spectrum of the narrow band burst, and the frequency-averaged polarisation profile of the burst, respectively. The PPA angles are shown for time bins with Stokes-I S/N$>$4. The linear polarisation, circular polarisation, and the Stokes-I frequency-averaged data are shown in red, blue, and black curves in Panel C. The RFI zapped channels are indicated with deep red colours. The burst arrival time in MJD is shown at the top of the figure.}\label{fig:ms_carbon_copies}
\end{figure*}

\begin{figure}
    \centering
    \begin{subfigure}[b]{1\textwidth}
        \centering
       \includegraphics[width=1\columnwidth]{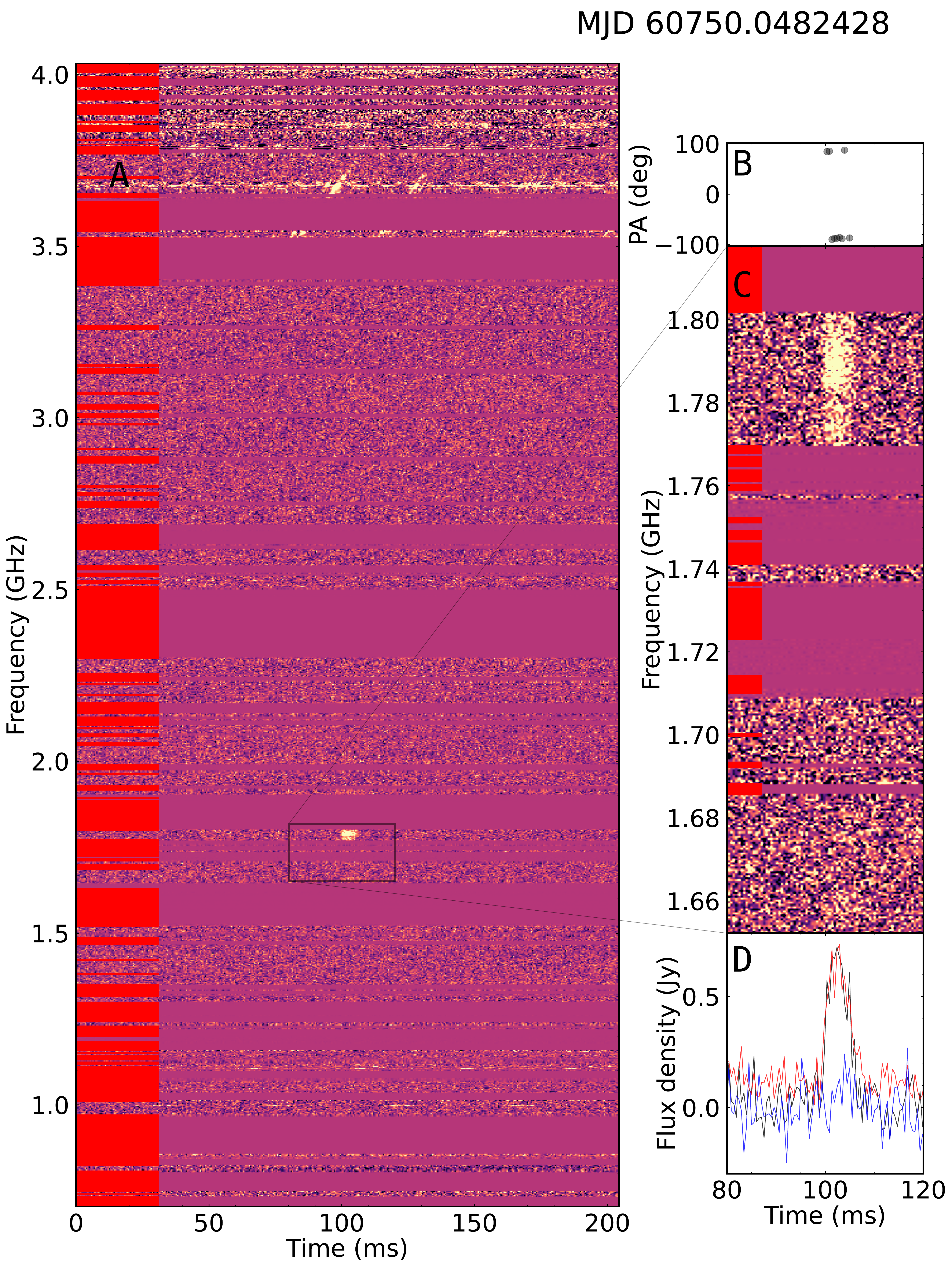}
        \caption{}
        \label{fig:narrow_carbon_copies_sub1}
    \end{subfigure}
    \caption{}
\end{figure}

\begin{figure}
    \ContinuedFloat
    \begin{subfigure}[b]{1\textwidth}
        \centering
        \includegraphics[width=1\columnwidth]{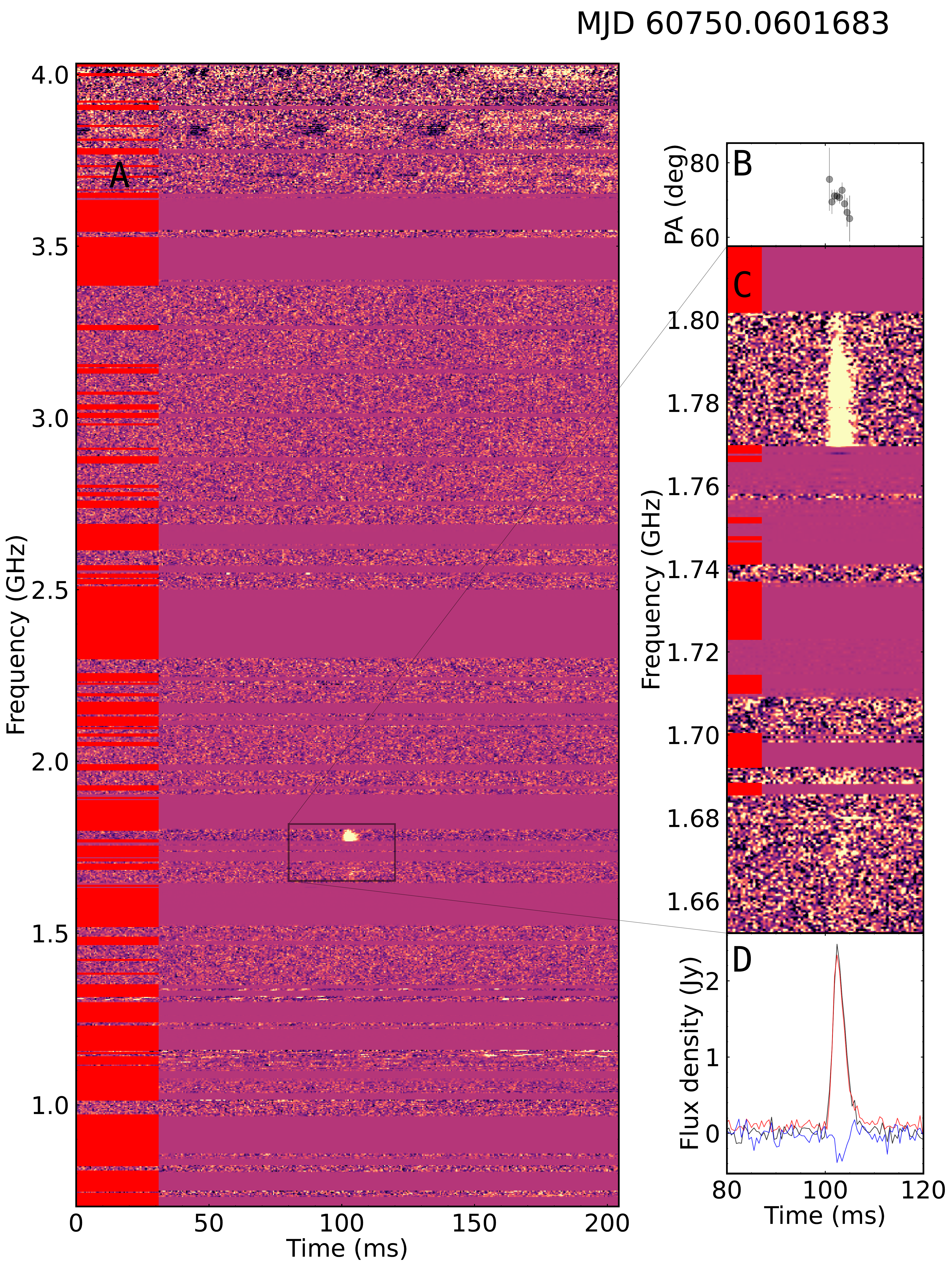}
        \caption{}
        \label{fig:narrow_carbon_copies_sub2}
    \end{subfigure}
    \caption{Continued}
\end{figure}

\begin{figure*}
    \ContinuedFloat
    \begin{subfigure}[b]{1\textwidth}
        \centering
        \includegraphics[width=1\columnwidth]{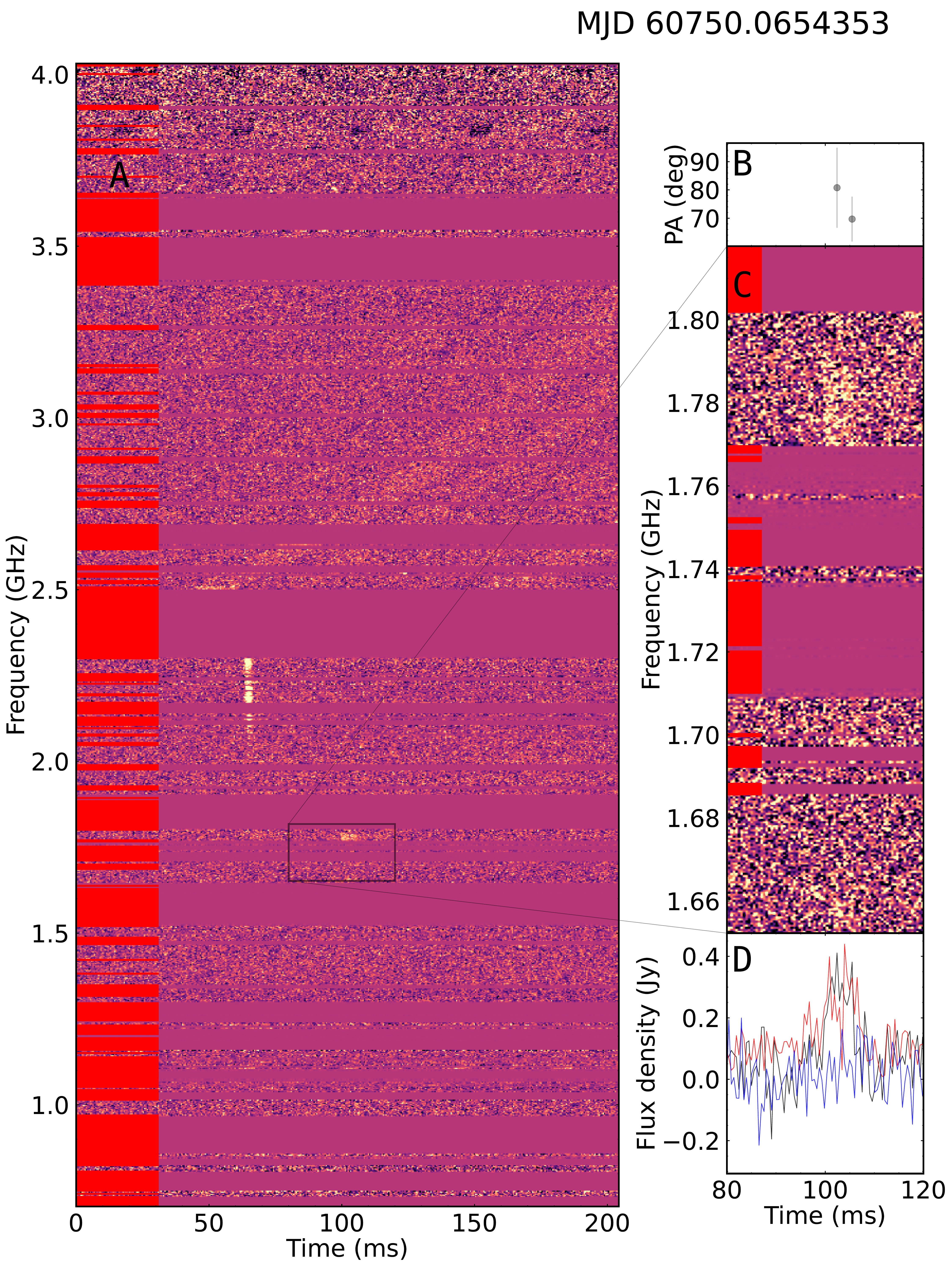}
        \caption{}
        \label{fig:narrow_carbon_copies_sub3}
    \end{subfigure}
    \caption{Continued}
\end{figure*}

\begin{figure*}[!htbp]
    \ContinuedFloat
    \caption{Narrow band carbon copies of bursts from MJD 60750 epoch. We show three bursts that share the common central frequencies and similar spectral morphologies. Relative to the burst shown in Figure \protect\ref{fig:narrow_carbon_copies}(a), the bursts in \protect\ref{fig:narrow_carbon_copies}(b) and (c) arrive 1030.37 and 1485.44 seconds later, respectively. The spectrum for the whole band and the inset plot have temporal and spectral resolution of 0.512 ms and 0.5 MHz, respectively. Panels A, B, C, and D show the dynamic spectrum of the burst with the whole band, the PPA angle, the zoomed-in spectrum of the narrow band burst, and the frequency-averaged polarisation profile of the burst. The PPA angles are shown for time bins that have Stokes-I S/N $>$4. The linear polarisation, circular polarisation, and the Stokes-I frequency-averaged data are shown in red, blue, and black curves in Panel C. The RFI zapped channels are indicated with deep red colours. The burst arrival times in MJD are shown for each of the subplots at the top of the figure.}\label{fig:narrow_carbon_copies}
\end{figure*}

\begin{figure*}
    \begin{subfigure}[b]{1\textwidth}
    \centering
        \includegraphics[width=1\columnwidth]{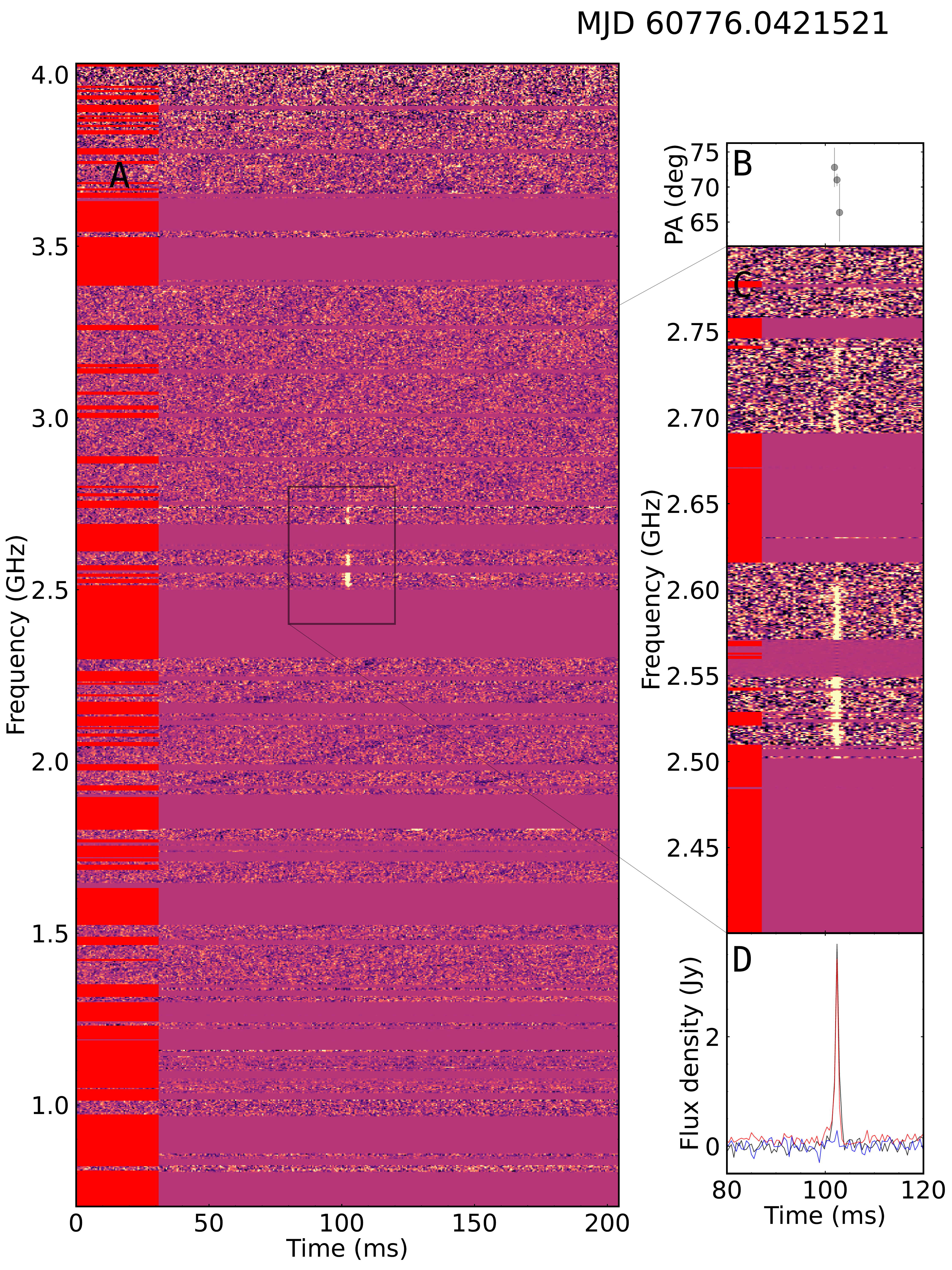}
        \caption{}\label{fig:narrow_carbon_copies1_sub1}
    \end{subfigure}
    \caption{}\label{}
\end{figure*}

\begin{figure*}
    \ContinuedFloat
    \begin{subfigure}[b]{1\textwidth}
    \centering
    \includegraphics[width=1\columnwidth]{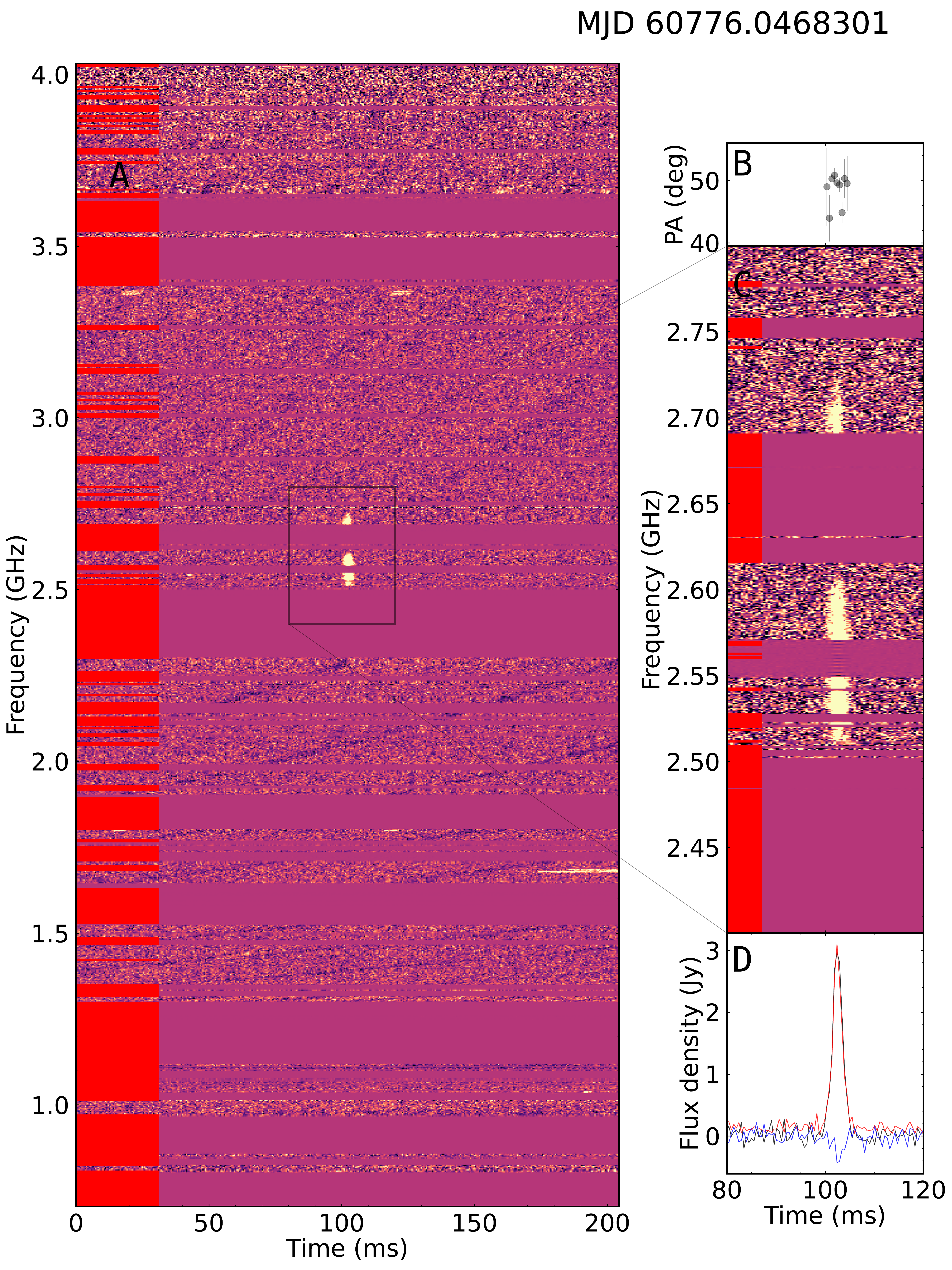}
    \caption{}\label{fig:narrow_carbon_copies1_sub2}
    \end{subfigure}    
    \caption{Continued}\label{}
\end{figure*}

\begin{figure*}[!htbp]
    \ContinuedFloat
    \caption{Narrow band carbon copies of bursts on MJD 60776 epoch. Figures show two bursts that share similar central frequencies and spectral morphologies. The burst arrival times are shown at the top of each figure. The two bursts were detected 404 seconds apart. The spectrum for the whole band and the inset plot have temporal and spectral resolution of 0.512 ms and 0.5 MHz, respectively. Panels A, B, C, and D show the burst with the entire frequency coverage of the UWL, the PPA angle, the zoomed-in spectrum of the narrow band burst, and the frequency-averaged polarisation profile of the burst, respectively. The PPA angles are shown for time bins with Stokes-I S/N$>$4. The linear polarisation, circular polarisation, and the Stokes-I frequency-averaged data are shown in red, blue, and black curves in Panel C, respectively. The RFI zapped channels are indicated with deep red colours.}\label{fig:narrow_carbon_copies1}
\end{figure*}

\begin{figure*}
    \includegraphics[width=0.51\columnwidth]{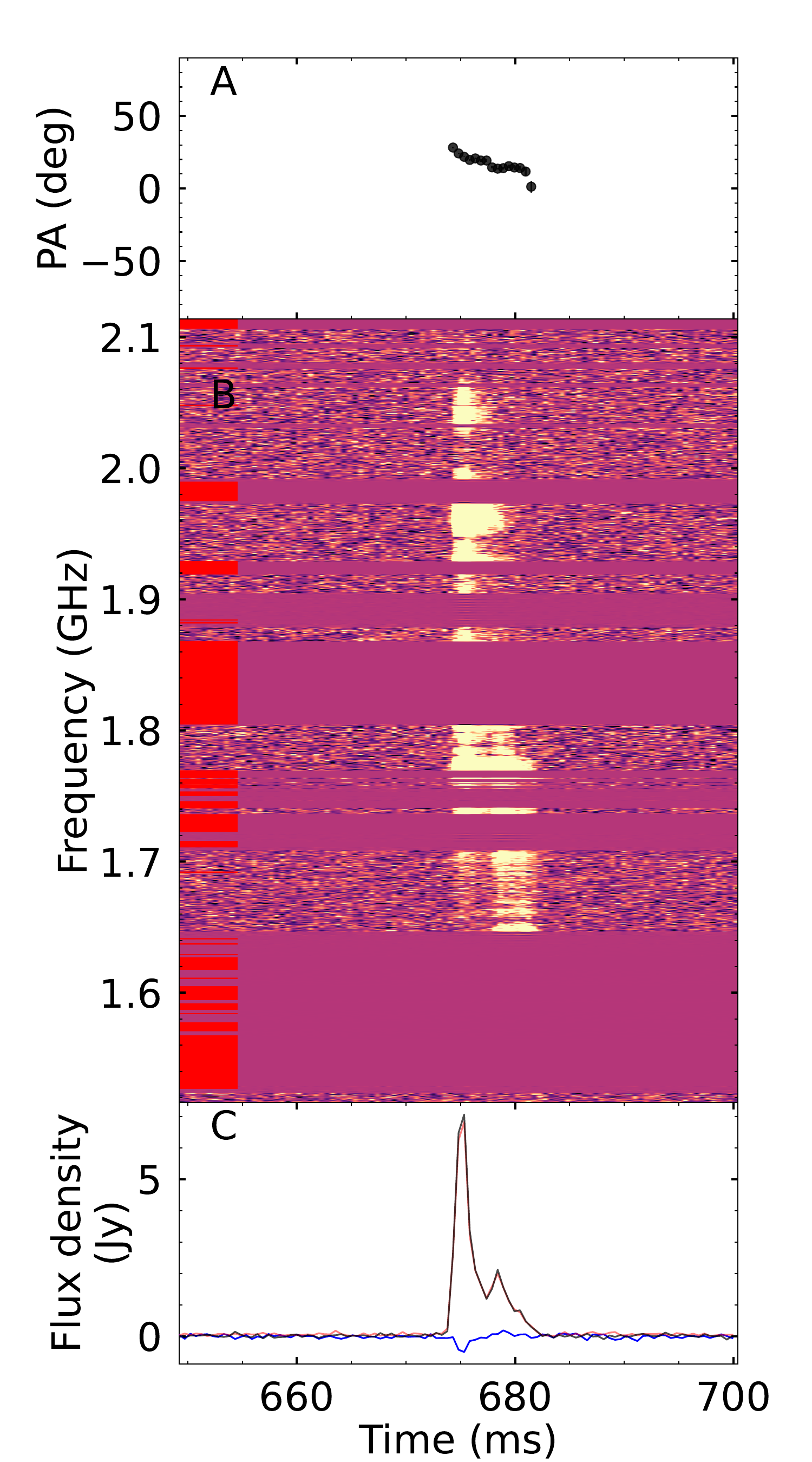}
    \includegraphics[width=0.51\columnwidth]{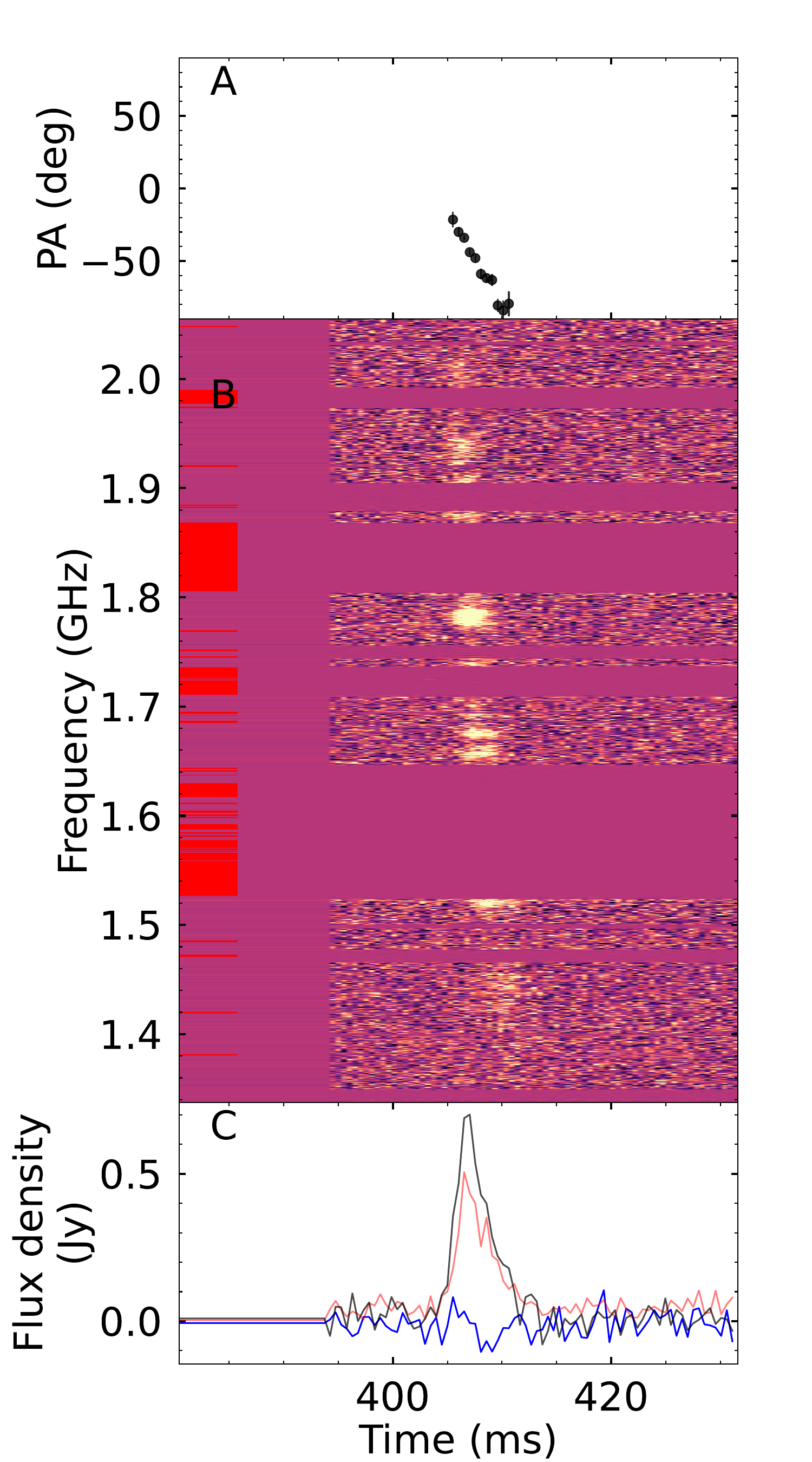}\\
    \caption{ Bursts showing sweep in the PPA angle. We show two repeat bursts from FRB 20240114A which show a PPA sweep across the on-pulse of the burst. Panels A, B, and C show the PPA angle, the dynamic Stokes-I spectrum, and the frequency-averaged polarisation profiles, respectively. Panel C shows the total intensity, linear and circular polarisation profiles in black, red, and blue, respectively. The PPA angles are shown for time bins with Stokes-I S/N$>$4. Figures are shown with a temporal and spectral resolution of 0.5 ms and 0.5 MHz, respectively. The RFI zapped channels are shown in deep red.}    \label{fig:PA_swing}
\end{figure*}

\begin{figure}
    \centering
    \vspace{2cm}
    \includegraphics[width=1.0\linewidth]{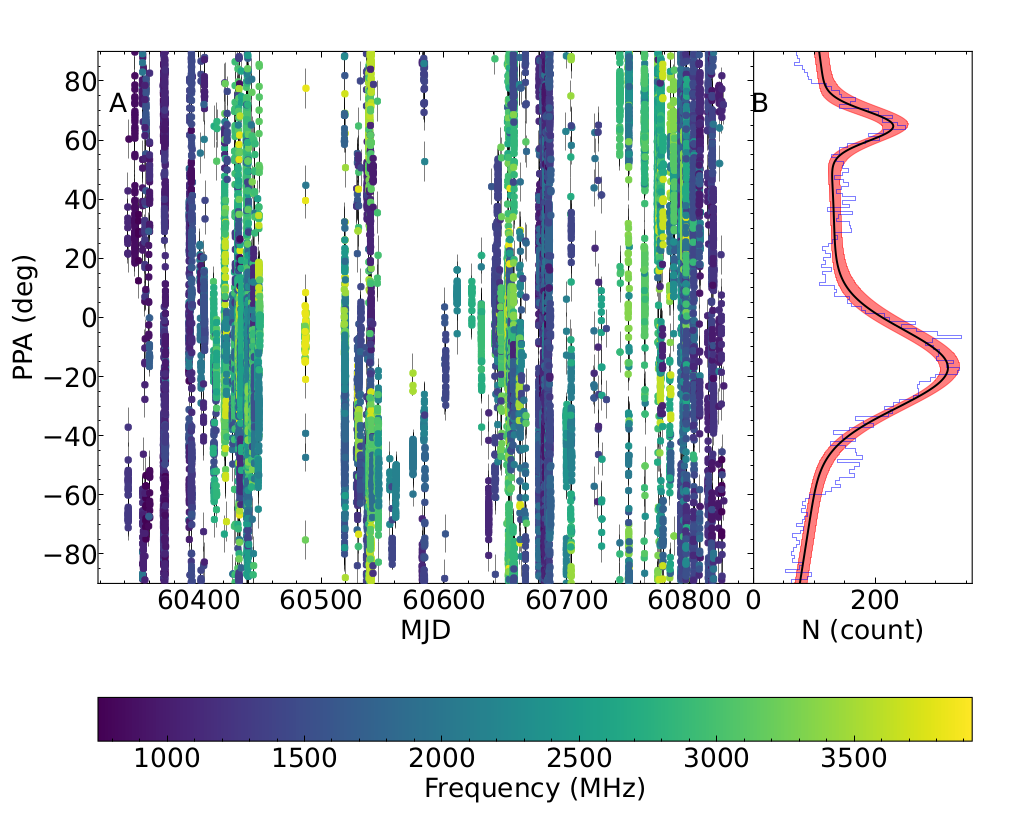}
    \caption{Polarisation Position Angle variations for FRB~20240114A. Panels A and B show the PPA evolution of the source through our campaign. Panel A shows PPA angles for bursts that were detected above an S/N of 20. Panel B shows the distribution of PPA angles with 150 bins in blue. We observe evidence of a multi-modal PPA distribution. The black curve shows the three-component Gaussian fit to the data. The red-shaded region shows the 1-$\sigma$ error region. The PPA angles from each burst are colour-coded with the centre frequency of the burst. 
    }
    \vspace{2cm}
    \label{fig:PA_evolution}
\end{figure}

\begin{figure}
    \vspace{2cm}
    \centering
    \includegraphics[width=1.0\linewidth]{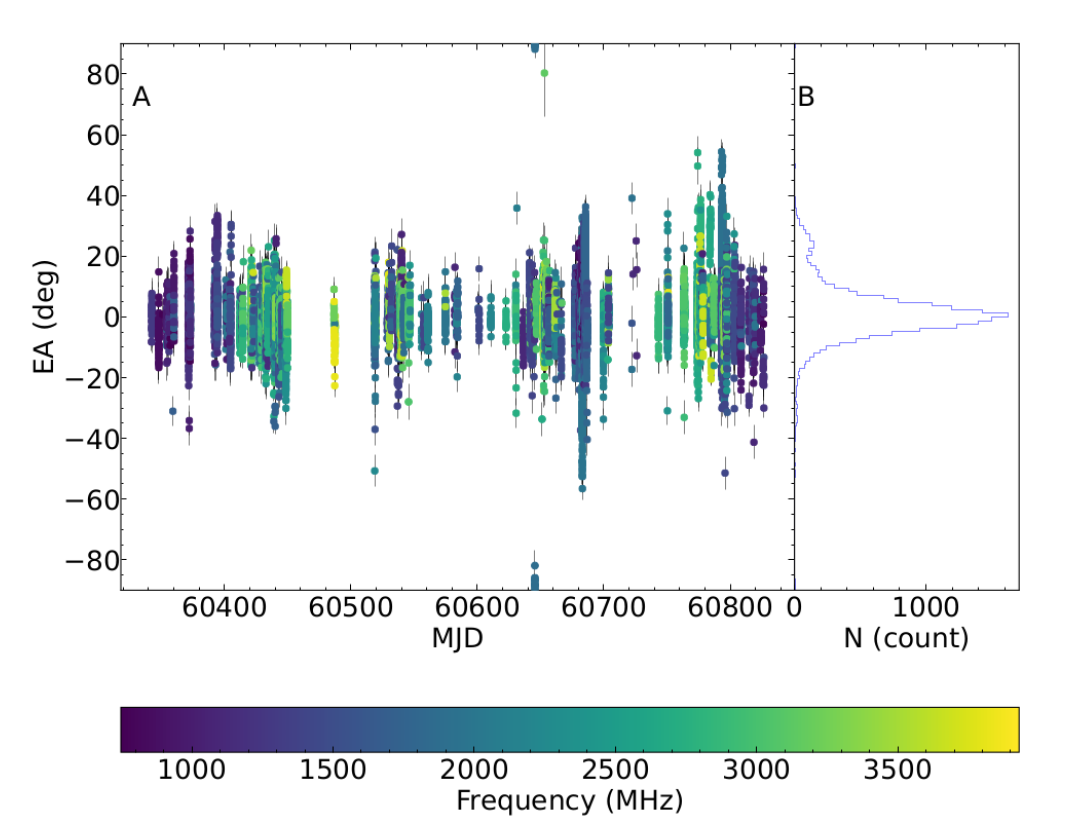}
    \caption{Ellipticity angle variations of FRB~20240114A. Panels A and B show the EA evolution of the source through the follow-up duration of $\sim$16 months. Panel A shows EA angles for bursts that were detected above an S/N of 20. Panel B shows the distribution of the EA angles with 150 bins in blue. The EA angles from each burst are colour-coded with the centre frequency of the bursts. 
    }
    \vspace{2cm}
    \label{fig:EA_evolution}
\end{figure}

\begin{figure}
    \centering
    \includegraphics[width=1\linewidth]{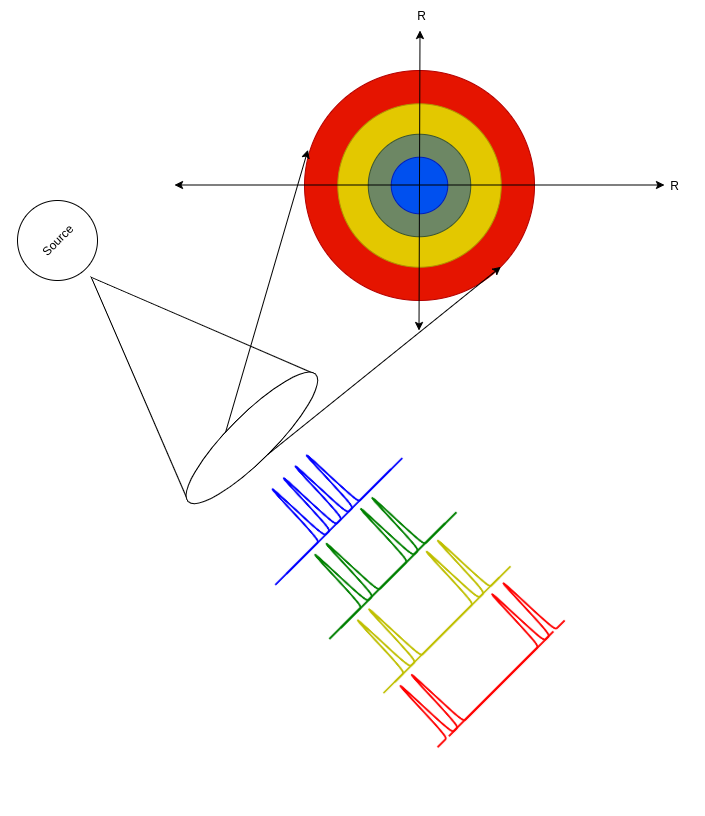}
    \caption{
    A schematic model for the structured jet scenario for FRB emission (adapted from Figure~5 of \citep{Sridhar+21}). Higher frequency bursts are launched close to the spine of the jet from a smaller radius, whereas lower frequency bursts are launched closer to the edge of the jet from a larger radius. The sweeping of the jet across the line of sight will lead to an evolution in the spectral emission. The red colour indicates lower-frequency emission, and the blue indicates higher-frequency emission.}
    \label{fig:intrinsic_toy}
\end{figure}

\begin{landscape}
\begin{flushleft}
\setlength{\tabcolsep}{1pt}
\renewcommand{\arraystretch}{1.7}
\setlength{\tabcolsep}{7pt}
\begin{longtable}{c c c c c c c c c c c}  
		\hline
		\thead{Epoch \\ (UTC)} & \thead{Epoch \\ (MJD)} & \thead{Bursts \\ detected \\ (N)} & \thead{On source time \\ (hrs)} & \thead{Burst rate \\ (\bursth)} & \thead{RM \\ (\RMunits)} & \thead{DM$_{\rm STRUCT}$ \\ (\DMunits)} & \thead{DM$_{\rm S/N}$ \\ (\DMunits)}& \thead{Bursts \\ (S/N $>$ 20)} & Receiver\\    
    \hline
    \endhead
			02-02-2024 & 60342 & 4 & 0.98 & 5.08$\pm$2.25 & 338.74$\pm$1.79 & 528.57$\pm$1.72 & 532.95$\pm$0.21 & 2 & UWL \\
			07-02-2024 & 60347 & 1 & 0.56 & 1.79$\pm$1.34 & 343.92$\pm$0.14 & 511.42$\pm$2.9 & 529.51$\pm$0.31 & 1 & UWL \\
			10-02-2024$^*$ & 60350 & 1 & 0.29 & 3.44$\pm$1.85 & 341.81$\pm$2.91 & - & 530.98$\pm$0.37 & - & UWL \\
			11-02-2024$^*\parallel$ & 60351 & 1 & 0.49 & 2.03$\pm$1.42 & - & - & - & - & UWL \\
			14-02-2024 & 60354 & 28 & 1.75 & 15.96$\pm$4.00 & 341.85$\pm$0.64 & 527.2$\pm$1.62 & 529.02$\pm$0.13 & 10 & UWL \\
			16-02-2024 & 60356 & 48 & 1.44 & 33.44$\pm$5.78 & 328.21$\pm$0.2 & 529.36$\pm$1.11 & 529.84$\pm$0.15 & 7 & UWL \\
			19-02-2024 & 60359 & 33 & 1.11 & 29.74$\pm$5.45 & 333.6$\pm$0.12 & 522.55$\pm$1.68 & 529.06$\pm$0.21 & 5 & UWL \\
			02-03-2024 & 60371 & 98 & 1.01 & 96.95$\pm$9.85 & 342.46$\pm$0.11 & 529.57$\pm$0.81 & 530.62$\pm$0.12 & 18 & UWL \\
			03-03-2024 & 60372 & 178 & 1.93 & 92.35$\pm$9.61 & 340.01$\pm$0.12 & 531.72$\pm$0.57 & 530.59$\pm$0.09 & 31 & UWL \\
			23-03-2024 & 60392 & 53 & 0.88 & 60.36$\pm$7.77 & 370.86$\pm$0.23 & 527.4$\pm$0.92 & 532.66$\pm$0.11 & 17 & UWL \\
			25-03-2024 & 60394 & 142 & 1.76 & 80.80$\pm$8.99 & 356.14$\pm$0.22 & 532.62$\pm$1.05 & 533.77$\pm$0.13 & 16 & UWL \\
			01-04-2024 & 60401 & 63 & 1.40 & 44.96$\pm$6.71 & 375.3$\pm$0.76 & 529.11$\pm$1.84 & 535.01$\pm$0.18 & 5 & UWL \\
			03-04-2024 & 60403 & 32 & 2.86 & 11.19$\pm$3.34 & 381.05$\pm$0.31 & 534.04$\pm$1.56 & 533.24$\pm$0.18 & 7 & UWL \\
			05-04-2024 & 60405 & 4 & 1.32 & 3.03$\pm$1.74 & 388.98$\pm$1.67 & 530.94$\pm$3.37 & 531.26$\pm$0.47 & 2 & UWL \\
			11-04-2024 & 60411 & 20 & 1.89 & 10.58$\pm$3.25 & 395.15$\pm$1.27 & 528.65$\pm$3.13 & 538.18$\pm$0.26 & - & UWL \\
			13-04-2024 & 60413 & 58 & 1.75 & 33.14$\pm$5.76 & 363.86$\pm$3.68 & 533.01$\pm$1.79 & 548.38$\pm$0.57 & - & UWL \\
			20-04-2024 & 60420 & 71 & 1.78 & 39.89$\pm$6.32 & 364.34$\pm$0.27 & 533.05$\pm$1.16 & 538.37$\pm$0.11 & 25 & UWL \\
			21-04-2024 & 60421 & 37 & 1.75 & 21.14$\pm$4.60 & 357.86$\pm$2.1 & 531.28$\pm$2.03 & 534.71$\pm$0.18 & 5 & UWL \\
			22-04-2024 & 60422 & 7 & 0.76 & 9.21$\pm$3.03 & 381.9$\pm$3.19 & 535.48$\pm$3.56 & 536.7$\pm$0.26 & - & UWL \\
			26-04-2024 & 60426 & 57 & 1.77 & 32.20$\pm$5.67 & 373.34$\pm$0.74 & 537.08$\pm$2.77 & 539.29$\pm$0.26 & 3 & UWL \\
			28-04-2024 & 60428 & 100 & 1.27 & 78.74$\pm$8.87 & 356.48$\pm$0.36 & 535.41$\pm$1.01 & 540.31$\pm$0.12 & 12 & UWL \\
			01-05-2024 & 60431 & 180 & 1.84 & 97.83$\pm$9.89 & 359.47$\pm$0.36 & 533.22$\pm$0.72 & 538.19$\pm$0.07 & 52 & UWL \\
			02-05-2024 & 60432 & 93 & 1.28 & 72.66$\pm$8.52 & 353.93$\pm$0.97 & 531.55$\pm$2.1 & 545.83$\pm$0.12 & 25 & UWL \\
			03-05-2024 & 60433 & 47 & 0.79 & 59.49$\pm$7.71 & 363.01$\pm$1.43 & 523.71$\pm$1.45 & 538.77$\pm$0.14 & 11 & UWL \\
			08-05-2024 & 60438 & 181 & 2.39 & 75.73$\pm$8.70 & 351.72$\pm$0.38 & 533.84$\pm$0.78 & 536.5$\pm$0.08 & 40 & UWL \\
			09-05-2024 & 60439 & 176 & 2.01 & 87.56$\pm$9.36 & 351.95$\pm$0.33 & 534.47$\pm$0.6 & 530.22$\pm$0.06 & 72 & UWL \\
			12-05-2024 & 60442 & 44 & 1.33 & 33.08$\pm$5.75 & 352.32$\pm$0.59 & 526.46$\pm$1.31 & 538.42$\pm$0.16 & 11 & UWL \\
			13-05-2024 & 60443 & 32 & 1.40 & 22.86$\pm$4.78 & 353.53$\pm$1.04 & 536.44$\pm$2.0 & 548.7$\pm$0.31 & 5 & UWL \\
			14-05-2024 & 60444 & 49 & 1.43 & 34.27$\pm$5.85 & 378.22$\pm$0.62 & 528.63$\pm$1.46 & 546.66$\pm$0.16 & 8 & UWL \\
			17-05-2024 & 60447 & 30 & 0.79 & 37.97$\pm$6.16 & 365.6$\pm$1.54 & 535.15$\pm$2.45 & 539.53$\pm$0.26 & 4 & UWL \\
			18-05-2024 & 60448 & 56 & 1.20 & 46.67$\pm$6.83 & 349.16$\pm$0.58 & 533.88$\pm$1.2 & 544.49$\pm$0.11 & 14 & UWL \\
			26-06-2024 & 60486 & 39 & 0.95 & 41.05$\pm$6.41 & 344.26$\pm$3.92 & 533.13$\pm$1.45 & 550.72$\pm$0.27 & 6 & UWL \\
			27-07-2024 & 60518 & 182 & 1.44 & 126.39$\pm$11.24 & 370.03$\pm$0.81 & 533.97$\pm$0.78 & 535.92$\pm$0.09 & 34 & UWL \\
			06-08-2024 & 60528 & 122 & 2.08 & 58.65$\pm$7.66 & 342.97$\pm$0.2 & 535.27$\pm$1.13 & 540.95$\pm$0.19 & 8 & UWL \\
			07-08-2024 & 60529 & 63 & 1.42 & 44.37$\pm$6.66 & 348.11$\pm$3.22 & 533.34$\pm$1.35 & 558.04$\pm$0.21 & 8 & UWL \\
			09-08-2024 & 60531 & 24 & 1.13 & 21.24$\pm$4.61 & 355.39$\pm$0.88 & 525.36$\pm$1.48 & 532.24$\pm$0.2 & 4 & UWL \\
			14-08-2024 & 60536 & 53 & 1.59 & 33.33$\pm$5.77 & 322.93$\pm$0.57 & 530.96$\pm$1.02 & 543.97$\pm$0.21 & 8 & UWL \\
			16-08-2024 & 60538 & 106 & 1.77 & 59.89$\pm$7.74 & 332.56$\pm$0.48 & 531.37$\pm$1.07 & 545.4$\pm$0.17 & 23 & UWL \\
			17-08-2024 & 60539 & 156 & 2.90 & 53.85$\pm$7.34 & 348.43$\pm$0.57 & 530.76$\pm$0.82 & 538.32$\pm$0.11 & 30 & UWL \\
			18-08-2024 & 60540 & 64 & 1.88 & 34.04$\pm$5.83 & 336.67$\pm$0.96 & 528.23$\pm$2.16 & 532.68$\pm$0.18 & 6 & UWL \\
			19-08-2024 & 60541 & 49 & 2 & 24.50$\pm$4.95 & 331.05$\pm$0.17 & 538.6$\pm$2.68 & 539.35$\pm$0.26 & 5 & UWL \\
			23-08-2024 & 60545 & 72 & 1.10 & 65.45$\pm$8.09 & 335.17$\pm$0.42 & 529.63$\pm$2.66 & 540.64$\pm$0.26 & 7 & UWL \\
			24-08-2024 & 60546 & 45 & 1.95 & 23.08$\pm$4.80 & 350.41$\pm$3.41 & 545.72$\pm$3.77 & 553.85$\pm$0.37 & 2 & UWL \\
			02-09-2024 & 60555 & 15 & 1.88 & 7.98$\pm$2.82 & 322.02$\pm$5.0 & 523.27$\pm$8.02 & 543.89$\pm$0.38 & - & UWL \\
			04-09-2024 & 60557 & 16 & 1.32 & 12.12$\pm$3.48 & 314.3$\pm$0.17 & 528.33$\pm$2.57 & 528.65$\pm$0.37 & 2 & UWL \\
			07-09-2024 & 60560 & 39 & 1.65 & 23.64$\pm$4.86 & 311.62$\pm$1.1 & 520.47$\pm$1.42 & 538.61$\pm$0.37 & 2 & UWL \\
			21-09-2024 & 60574 & 34 & 1.85 & 18.38$\pm$4.29 & 353.64$\pm$0.78 & 522.18$\pm$2.19 & 538.24$\pm$0.26 & 2 & UWL \\
			28-09-2024 & 60581 & 13 & 0.79 & 16.46$\pm$4.06 & 363.5$\pm$0.12 & 529.83$\pm$2.35 & 532.71$\pm$0.37 & 2 & UWL \\
			30-09-2024 & 60583 & 47 & 1.98 & 23.74$\pm$4.87 & 351.17$\pm$0.89 & 531.72$\pm$1.06 & 537.62$\pm$0.24 & 9 & UWL \\
			17-10-2024 & 60600 & 15 & 1.95 & 7.69$\pm$2.77 & 345.68$\pm$0.73 & 540.12$\pm$2.26 & 542.54$\pm$0.33 & 2 & UWL \\
			27-10-2024$^*$ & 60610 & 4 & 0.32 & 12.50$\pm$3.54 & 381.32$\pm$0.02 & - & 535.29$\pm$0.37 & 2 & UWL \\
			08-11-2024 & 60622 & 21 & 0.54 & 38.89$\pm$6.24 & 388.84$\pm$0.71 & 529.63$\pm$3.19 & 529.14$\pm$0.37 & 2 & UWL \\
			16-11-2024 & 60630 & 21 & 1.40 & 15.00$\pm$3.87 & 329.27$\pm$1.31 & 528.24$\pm$3.41 & 531.97$\pm$0.48 & 3 & UWL \\
			24-11-2024 & 60638 & 24 & 1.35 & 17.78$\pm$4.22 & 374.58$\pm$1.55 & 545.97$\pm$2.05 & 546.48$\pm$0.6 & 2 & UWL \\
			27-11-2024 & 60641 & 48 & 1.42 & 33.80$\pm$5.81 & 388.18$\pm$0.39 & 531.11$\pm$1.96 & 531.52$\pm$0.17 & 7 & UWL \\
			29-11-2024 & 60643 & 15 & 0.73 & 20.55$\pm$4.53 & 346.87$\pm$1.17 & 534.56$\pm$2.31 & 540.86$\pm$0.22 & 5 & UWL \\
			30-11-2024 & 60644 & 13 & 1.41 & 9.22$\pm$3.04 & 354.2$\pm$5.64 & 531.19$\pm$1.75 & 532.95$\pm$0.37 & 2 & UWL \\
			01-12-2024 & 60645 & 23 & 1.46 & 15.75$\pm$3.97 & 363.37$\pm$0.16 & 526.18$\pm$2.24 & 533.08$\pm$0.37 & 2 & UWL \\
			02-12-2024 & 60646 & 30 & 1.41 & 21.28$\pm$4.61 & 349.06$\pm$1.08 & 520.51$\pm$2.04 & 542.15$\pm$0.29 & 7 & UWL \\
			06-12-2024 & 60650 & 0 & 0.17 & - & - & - & - & - & MARS \\
			07-12-2024 & 60651 & 21 & 0.55 & 38.18$\pm$6.18 & 372.03$\pm$0.4 & 529.73$\pm$2.05 & 541.85$\pm$0.15 & 10 & UWL \\
			08-12-2024 & 60652 & 100 & 1.40 & 71.43$\pm$8.45 & 358.97$\pm$1.11 & 534.04$\pm$0.94 & 540.33$\pm$0.1 & 31 & UWL \\
			09-12-2024 & 60653 & 75 & 1.36 & 55.15$\pm$7.43 & 347.04$\pm$1.86 & 531.81$\pm$0.99 & 542.12$\pm$0.13 & 12 & UWL \\
			10-12-2024 & 60654 & 81 & 1.42 & 57.04$\pm$7.55 & 370.2$\pm$0.4 & 534.51$\pm$0.9 & 540.53$\pm$0.1 & 19 & UWL \\
			11-12-2024 & 60655 & 0 & 0.77 & - & - & - & - & - & MARS \\
			12-12-2024 & 60656 & 62 & 1.43 & 43.36$\pm$6.58 & 357.1$\pm$0.56 & 530.14$\pm$1.27 & 534.03$\pm$0.13 & 11 & UWL \\
			15-12-2024 & 60659 & 0 & 0.90 & - & - & - & - & - & MARS \\
			16-12-2024 & 60660 & 0 & 1.20 & - & - & - & - & - & MARS \\
			17-12-2024 & 60661 & 73 & 1.37 & 53.28$\pm$7.30 & 375.0$\pm$0.67 & 536.11$\pm$1.07 & 537.9$\pm$0.16 & 15 & UWL \\
			18-12-2024 & 60662 & 5 & 0.42 & 11.90$\pm$3.45 & 305.68$\pm$2.0 & 522.98$\pm$3.37 & 537.12$\pm$0.29 & 2 & UWL \\
			20-12-2024 & 60664 & 4 & 1.26 & 3.17$\pm$1.78 & 317.03$\pm$7.14 & 520.44$\pm$2.66 & 535.82$\pm$0.52 & 2 & UWL \\
			21-12-2024 & 60665 & 3 & 1.50 & 2.00$\pm$1.41 & 323.19$\pm$1.46 & 531.36$\pm$3.8 & 527.79$\pm$0.44 & 2 & UWL \\
			22-12-2024 & 60666 & 9 & 1.38 & 6.52$\pm$2.55 & 358.04$\pm$0.46 & 530.56$\pm$1.25 & 534.92$\pm$0.29 & 2 & UWL \\
			02-01-2025 & 60677 & 130 & 2.10 & 61.90$\pm$7.87 & 342.49$\pm$0.24 & 531.05$\pm$1.59 & 528.66$\pm$0.08 & 24 & UWL \\
			05-01-2025 & 60680 & 34 & 1.30 & 26.15$\pm$5.11 & 341.9$\pm$0.11 & 530.57$\pm$0.75 & 532.42$\pm$0.12 & 10 & UWL \\
			06-01-2025 & 60681 & 33 & 1.24 & 26.61$\pm$5.16 & 336.71$\pm$0.13 & 528.9$\pm$1.18 & 531.93$\pm$0.09 & 17 & UWL \\
			07-01-2025 & 60682 & 23 & 1.37 & 16.79$\pm$4.10 & 327.64$\pm$0.16 & 528.65$\pm$1.27 & 534.77$\pm$0.21 & 10 & UWL \\
			09-01-2025 & 60684 & 75 & 1.46 & 51.37$\pm$7.17 & 331.5$\pm$0.2 & 530.19$\pm$3.25 & 535.12$\pm$0.12 & 24 & UWL \\
			10-01-2025 & 60685 & 88 & 1.25 & 70.40$\pm$8.39 & 341.96$\pm$0.16 & 528.07$\pm$0.68 & 533.21$\pm$0.12 & 17 & UWL \\
			11-01-2025 & 60686 & 44 & 0.86 & 51.16$\pm$7.15 & 346.05$\pm$0.36 & 524.73$\pm$1.73 & 531.46$\pm$0.14 & 8 & UWL \\
			24-01-2025 & 60699 & 44 & 1.45 & 30.34$\pm$5.51 & 335.02$\pm$0.51 & 532.72$\pm$1.39 & 536.83$\pm$0.15 & 7 & UWL \\
			28-01-2025 & 60703 & 24 & 1.44 & 16.67$\pm$4.08 & 310.03$\pm$0.59 & 531.28$\pm$1.77 & 537.68$\pm$0.17 & 8 & UWL \\
			03-02-2025 & 60709 & 0 & 0.87 & - & - & - & - & - & MARS \\
			16-02-2025 & 60722 & 4 & 1 & 4.00$\pm$2.00 & 285.57$\pm$1.26 & 530.5$\pm$3.45 & 530.5$\pm$3.45 & 2 & UWL \\
			19-02-2025$^*\parallel$ & 60725 & 3 & 1.40 & 2.14$\pm$1.46 & - & - & - & 2 & UWL \\
			22-02-2025$^*\parallel$ & 60728 & 2 & 1.02 & 1.96$\pm$1.40 & - & - & - & - & UWL \\
			26-02-2025$^*\parallel$ & 60732 & 2 & 0.94 & 2.13$\pm$1.46 & - & - & - & - & UWL \\
			09-03-2025 & 60743 & 7 & 0.45 & 15.56$\pm$3.94 & 293.29$\pm$4.28 & 516.99$\pm$2.43 & 555.09$\pm$0.31 & 2 & UWL \\
			16-03-2025 & 60750 & 83 & 2.91 & 28.52$\pm$5.34 & 300.5$\pm$2.28 & 525.07$\pm$1.35 & 535.21$\pm$0.13 & 19 & UWL \\
			21-03-2025$^*\parallel$ & 60755 & 8 & 0.50 & 16.00$\pm$4.00 & - & - & - & - & UWL \\
			29-03-2025 & 60763 & 56 & 1.09 & 51.38$\pm$7.17 & 251.2$\pm$0.83 & 535.24$\pm$1.01 & 546.2$\pm$0.11 & 20 & UWL \\
			08-04-2025 & 60773 & 91 & 1.38 & 65.94$\pm$8.12 & 259.6$\pm$0.61 & 529.66$\pm$1.09 & 544.06$\pm$0.09 & 28 & UWL \\
			09-04-2025 & 60774 & 28 & 0.52 & 53.85$\pm$7.34 & 271.58$\pm$1.03 & 529.43$\pm$1.44 & 538.95$\pm$0.19 & 13 & UWL \\
			11-04-2025 & 60776 & 55 & 0.91 & 60.44$\pm$7.77 & 248.23$\pm$0.76 & 525.84$\pm$1.11 & 544.52$\pm$0.11 & 27 & UWL \\
			12-04-2025 & 60777 & 18 & 1.34 & 13.43$\pm$3.67 & 264.31$\pm$2.3 & 533.37$\pm$1.58 & 538.24$\pm$0.19 & 5 & UWL \\
			18-04-2025 & 60783 & 25 & 0.90 & 27.78$\pm$5.27 & 250.86$\pm$1.17 & 529.52$\pm$1.12 & 537.45$\pm$0.14 & 11 & UWL \\
			21-04-2025 & 60786 & 45 & 3.14 & 14.33$\pm$3.79 & 227.53$\pm$0.63 & 534.99$\pm$1.49 & 541.34$\pm$0.14 & 11 & UWL \\
			27-04-2025 & 60792 & 67 & 1.10 & 60.91$\pm$7.80 & 224.34$\pm$0.7 & 533.16$\pm$0.87 & 538.69$\pm$0.09 & 30 & UWL \\
			29-04-2025 & 60794 & 97 & 1.42 & 68.31$\pm$8.26 & 227.75$\pm$0.36 & 532.28$\pm$0.63 & 537.61$\pm$0.07 & 46 & UWL \\
			01-05-2025 & 60796 & 88 & 1.55 & 56.77$\pm$7.53 & 220.56$\pm$0.43 & 531.3$\pm$1.64 & 526.75$\pm$0.07 & 40 & UWL \\
			07-05-2025 & 60802 & 239 & 2.76 & 86.59$\pm$9.31 & 222.6$\pm$0.26 & 529.33$\pm$0.6 & 531.51$\pm$0.06 & 51 & UWL \\
			12-05-2025 & 60807 & 74 & 2.90 & 25.52$\pm$5.05 & 256.56$\pm$0.34 & 527.88$\pm$0.71 & 531.95$\pm$0.1 & 18 & UWL \\
			19-05-2025 & 60814 & 54 & 1.39 & 38.85$\pm$6.23 & 240.66$\pm$0.18 & 530.61$\pm$1.05 & 531.84$\pm$0.16 & 11 & UWL \\
			22-05-2025 & 60817 & 69 & 1.92 & 35.94$\pm$5.99 & 246.8$\pm$0.2 & 529.75$\pm$0.8 & 532.41$\pm$0.11 & 16 & UWL \\
			23-05-2025 & 60818 & 28 & 1.42 & 19.72$\pm$4.44 & 255.2$\pm$0.1 & 530.12$\pm$0.93 & 532.92$\pm$0.23 & 7 & UWL \\
			26-05-2025 & 60821 & 52 & 2.29 & 22.71$\pm$4.77 & 255.91$\pm$0.5 & 531.67$\pm$1.85 & 530.95$\pm$0.28 & 6 & UWL \\
			28-05-2025 & 60823 & 34 & 1.46 & 23.29$\pm$4.83 & 232.69$\pm$0.16 & 531.09$\pm$1.27 & 530.92$\pm$0.16 & 8 & UWL \\
			30-05-2025 & 60825 & 29 & 1.29 & 22.48$\pm$4.74 & 248.34$\pm$0.19 & 530.89$\pm$1.35 & 531.2$\pm$0.13 & 9 & UWL \\
			01-06-2025 & 60827 & 15 & 0.97 & 15.46$\pm$3.93 & 240.58$\pm$0.22 & 531.94$\pm$1.96 & 531.27$\pm$0.21 & 3 & UWL \\
            \bottomrule
            \caption{The burst activity from FRB 20240114A. The table lists the total number of detected bursts from UWL for each epoch. The MJD of each epoch corresponds to the day of the observation. The burst rate is calculated using the on-source time. The RM for all bursts with S/N$>$20 is calculated using 1D RM Synthesis \code{rmtools}\protect\footnote{\url{https://github.com/CIRADA-Tools/RM-Tools}}. $^{\dag}$The epochs which do not have bursts S/N $>$ 20, we use lower significance bursts to estimate RMs. All the bursts are identified as true bursts from human vetting of the candidates obtained from \code{fetch} \protect\citep{Agarwal:2019} and manually checking all the candidates between the DM range of 520 and 540 \DMunits. We report an average DM for each epoch using bursts that have a detection S/N of $>$20. We use \code{DMphase}\protect\cite{Seymour:2019} to estimate the structure maximised DM and \code{pdmp} for S/N maximisation\protect\cite{vanStraten:2010}. One-$\sigma$ uncertainties are reported. The uncertainties for DM and RM are estimated using propagation of uncertainties. \\$^\dag$No polarisation information available because of the corruption of the noise diode data.
            \\$^*$We do not report DM$_{\rm struct}$ due to low S/N bursts or broadband affected channels.
            \\$^\parallel$ No RM measurement due to low S/N bursts.}

    \label{tab:epoch_list}
\end{longtable}
\end{flushleft}
\end{landscape}
\begin{center}
    \setlength{\tabcolsep}{7pt}
    \renewcommand{\arraystretch}{2}
  \begin{longtable}{c c c c c c}
		\toprule
			 Burst storm & $\alpha$ & v$_{\rm trans}$ & t$_{\rm tansshift}$ & $\mathcal{G}$ & MJD\\
			\midrule
            \endhead
			B4 & 0.9$^{+0.1}_{-0.0}$ & 7.9$^{+0.2}_{-0.2}$ & 1.2$^{+0.3}_{-0.3}$ & 5.6$^{+0.1}_{-0.1}$ & 60610 - 60725 \\
			B5 & 0.6$^{+0.1}_{-0.1}$ & 8.0$^{+0.6}_{-0.4}$ & -4.1$^{+0.4}_{-0.4}$ & 7.6$^{+0.2}_{-0.2}$ & 60750 - 60825 \\
            \hline
            \caption{Lens parameters from Gaussian lens fit. The table lists the lens parameters from the Gaussian lens model to the burst rate. We denote the two burst storms after 60600 MJD as B4 and B5. The $\alpha$ value encapsulates the physical parameters DM$_{l}$, frequency ($\nu$), the lens scale $a$ and d$_{\rm sl}$. The v$_{\rm trans}$ factor is the arbitrary scaling unit used to fit the data. We scale the x-axis as we do not know the transverse velocity of the source \protect\cite{Clegg:1998}. The t$_{\rm transshift}$ is the factor used to find the minima of the burst activity. $\mathcal{G}$ is the baseline gain.}\label{tab:len_params}
    \end{longtable}
\end{center}

\begin{center}
    \renewcommand{\arraystretch}{1.5}
    \setlength{\tabcolsep}{4pt}
    \begin{longtable}{c c c c c c c}
		\toprule
			Epoch & MJD & \thead{Detection DM \\ (\DMunits)} & \thead{DM$_{\rm STRUCT}$ \\ (\DMunits)} & \thead{\thead{Bandwidth \\ (MHz)}} & \thead{Centre Frequency \\ (MHz)} & ToA \\
		\midrule
            2024-12-02        & 60646 & 549.36 & 546.24$\pm$0.37 & 340 & 3025 & 60646.178782 \\
                              &       & 535.01 & 535.29$\pm$0.37 & 340 & 3025 & 60646.179538 \\
            \hline
            2025-03-16	      & 60750 & 539.51 & 539.22$\pm$0.37 & 150 & 1735 & 60750.048243 \\
			                 &       & 534.38 & 533.81$\pm$0.37 & 150 & 1735 & 60750.056335 \\
			                 &       & 534.79 & 537.0$\pm$0.37 & 150 & 1735 & 60750.060168 \\
			                 &       & 533.46 & 540.32$\pm$0.87 & 150 & 1735 & 60750.065435 \\
			                 &       & 529.54 & 526.68$\pm$ 0.42 & 150 & 1735 & 60750.094312 \\
            \hline
            2025-04-11        & 60776 & 528.73 & 529.4619$\pm$0.04& 223 & 2600 & 60776.042152 \\
                              &       & 535.49 & 536.57$\pm$0.37& 223 & 2600 & 60776.045730 \\
                              &       & 533.55 & 539.35$\pm$0.37 & 223 & 2600 & 60776.046830 \\

            \bottomrule
            \caption{Properties of bursts with spectral memory. We report spectral memory bursts from three epochs that were separated by timescales of minutes. We report the search pipeline detection DM, the structure maximised DM using PDMP, bandwidth, and central frequency for each spectral memory burst. We show example spectral memory bursts in Extended Data Figures \ref{fig:narrow_carbon_copies} and \ref{fig:narrow_carbon_copies1}. We visually extract the on-pulse region of the bursts. The centre frequencies are reported for these extracted archives. We also report ToAs from \code{heimdall}, which are associated with the highest S/N candidate in the subbanded search.}\label{tab:spec_mem}
    \end{longtable}
\end{center}






\clearpage

\bibliography{sn-bibliography}

\end{document}